\documentclass[11pt]{article}
\pdfoutput=1
\usepackage[dvipdfmx]{graphicx}
\usepackage{jheppub}
\usepackage{braket} 
\usepackage{url}

\definecolor{refkey}{rgb}{0.9451,0.2706,0.4941}
\definecolor{labelkey}{rgb}{0.9451,0.2706,0.4941}

\usepackage{tikz}
\usetikzlibrary{intersections, calc,positioning,decorations.pathreplacing,decorations.pathmorphing}

\usepackage{amsthm}
\usepackage{comment}

\usepackage{bm}
\usepackage[utf8]{inputenc}
\usepackage{amsmath}
\usepackage{slashed}

\def\d{{\rm d}}
\def\i{{\rm i}}
\def\CA{{\cal A}}

\def\CJ{{\cal J}}
\def\CI{{\cal I}}

\def\CM{{\cal M}}
\def\CN{{\cal N}}
\def\CO{{\cal O}}

\def\CW{{\cal W}}

\def\BR{\mathbb{R}}
\def\BS{\mathbb{S}}

\def\d{\mathrm{d}}

\def\be{\begin{equation}}
\def\ee{\end{equation}}
\def\ba{\begin{eqnarray}}
\def\ea{\end{eqnarray}}
\def\bal#1\eal{\begin{align}#1\end{align}}

\newcommand{\del}{\partial} 
\newcommand{\dal}{\square} 
\newcommand{\sla}[1]{{\ooalign{\hfil/\hfil\crcr$#1$}}} 
\renewcommand{\bar}[1]{\overline{#1}} 
\newcommand{\abs}[1]{\left|#1\right|} 
\renewcommand{\Lambda}{\varLambda} 

\newcommand{\bbar}[1]{\bar{#1}{}}

\newcommand{\s}{\sigma}

\newcommand{\dlt}{\delta}
\newcommand{\e}{\epsilon}

\newcommand{\ve}{\varepsilon}

\theoremstyle{definition}

\title{
Janus interface entropy and Calabi's diastasis in four-dimensional $\mathcal{N}=2$ superconformal field theories}

\author[a,b]{Kanato Goto,}
\author[c]{Lento Nagano,}
\author[d]{Tatsuma Nishioka}
\author[c]{and Takuya Okuda}

\affiliation[a]{Department of Physics, Cornell University, Ithaca, New York, USA}
\affiliation[b]{RIKEN Interdisciplinary Theoretical and Mathematical Sciences (iTHEMS),\\
Wako, Saitama 351-0198, Japan}
\affiliation[c]{Institute of Physics, University of Tokyo, Komaba, \\ Meguro-ku, Tokyo 153-8902, Japan}
\affiliation[d]{Department of Physics, Faculty of Science,
The University of Tokyo,\\
Bunkyo-ku, Tokyo 113-0033, Japan}

\emailAdd{kanato.goto@riken.jp}
\emailAdd{nagano@hep1.c.u-tokyo.ac.jp}
\emailAdd{nishioka@hep-th.phys.s.u-tokyo.ac.jp}
\emailAdd{takuya@hep1.c.u-tokyo.ac.jp}
		
\abstract{%
We study the entropy associated with the Janus interface in a 4$d$ $\mathcal{N}=2$ superconformal field theory.
With the entropy defined as the interface contribution to an entanglement entropy we show, under mild assumptions, that the Janus interface entropy is proportional to the geometric quantity called Calabi's diastasis on the space of $\mathcal{N}=2$ marginal couplings, confirming an earlier conjecture by two of the authors and generalizing a similar result in two dimensions.
Our method is based on a CFT consideration that makes use of the Casini--Huerta--Myers conformal map from the flat space to the round sphere.
}

\preprint{RIKEN-iTHEMS-Report-20, UT-Komaba-20-1}

\begin{document}
\maketitle

\section{Introduction}


Interfaces in a quantum field theory are codimension-one objects that connect two neighboring regions in spacetime.
Though they exhibit rich physical properties, they have been as yet only partially explored.
Interfaces appear in various physical contexts such as condensed matter physics, supersymmetric field theories, and string theory.
In this paper we are particularly interested in the interfaces that are characterized by a spatial change in the values of the coupling constants; such interfaces are called Janus interfaces~\cite{Bak:2003jk,Clark:2004sb}.
More specifically, we will study the entanglement entropy associated with the Janus interface in four-dimensional (4$d$) $\mathcal{N}=2$ supersymmetric gauge theories.

For 2$d$ $\mathcal{N}=(2,2)$ superconformal field theories (SCFTs), the reference~\cite{Bachas:2013nxa} found an intriguing relation between the interface entropy (the $g$-function~\cite{Affleck:1991tk}) and the quantity~${\cal D}
$ known as Calabi's diastasis.
Let us consider the K\"ahler potential on the (super)conformal manifold, {\it i.e.}, the space of exactly marginal couplings~$\tau=(\tau_I)$ preserving $\mathcal{N}=(2,2)$ superconformal symmetry.
For notational simplicity and without loss of generality we assume that there is only one complex marginal coupling~$\tau$.
Let $\tau^*$ be the complex conjugate of $\tau$, and $\bar\tau$ an independent complex variable.
For $\bar\tau-\tau^*$ small enough, one can analytically continue the K\"ahler potential so that the function $K(\tau,\bar{\tau})$ that depends holomorphically on~$\tau$ and~$\bar{\tau}$ reduces to it when~$\bar{\tau}=\tau^*$~\cite{Calabi}.
Let $\tau_+$ and $\tau_-$ be two points that are close enough.
Calabi's diastasis is the function given by the following combination of the analytically continued K\"ahler potentials:
\ba\label{diastasis-def}
 {\cal D}
 :=K(\tau_+,\bar{\tau}_+)+K(\tau_-,\bar{\tau}_-)-K(\tau_+,\bar{\tau}_-)-K(\tau_-,\bar{\tau}_+)\, .
\ea 
It can be viewed as a measure of separation between the two points on the conformal manifold; it becomes proportional to the usual metric when the two points are infinitesimally close.
The finding of~\cite{Bachas:2013nxa} is that the $g$-function of the interface across which the couplings of the SCFT take different values $(\tau_+,\bar{\tau}_+)$ and $(\tau_-,\bar{\tau}_-)$ is given in terms of Calabi's diastasis function $ {\cal D}
$ as
 \ba\label{eq:g-factor-diastasis}
2\log g=  {\cal D}
\, .
\ea
This formula provides an interpretation of the interface entropy in terms of the geometry of the space of quantum field theories.
The claim of~\cite{Bachas:2013nxa} was further confirmed via holography~\cite{DHoker:2014qtw}, super-Weyl anomaly~\cite{Bachas:2016bzn},  and supersymmetric (SUSY) localization~\cite{Goto:2018bci}. 

A generalization of the relation~(\ref{eq:g-factor-diastasis}) to 4$d$ $\mathcal{N}=2$ theories was conjectured in~\cite{Goto:2018zrp}.
In general one can define the entropy $S_{\mathcal{I}}$ of an interface that separates CFT$_+$ and  CFT$_-$ as
\ba\label{interface-entropy-def}
S_{\mathcal{I}}=
S^\text{(ICFT)}_E 
-\frac{1}{2} \left(S_E^{(\text{CFT}_+)} +S_E^{(\text{CFT}_-)}\right) \,,
\ea
 where $S^\text{(ICFT)}_E $ is the entanglement entropy for a spherical entangling surface in the interface CFT (ICFT),
  and $S_E^{(\text{CFT}_{ \pm})}$ is the entanglement entropy computed using the same geometry for CFT$_\pm$ without an interface.
The reference~\cite{Goto:2018zrp} conjectured that the interface entropy $S_{\mathcal{I}}$ for a half-BPS Janus interface in a 4$d$ $\mathcal{N}=2$ SCFT is again proportional to Calabi's diastasis on the $\mathcal{N}=2$ conformal manifold
\ba
S_{\mathcal{I}} \propto {\cal D}
\, .\label{Conj}
\ea
In~\cite{Goto:2018zrp} the conjecture was confirmed for a special case, namely the large-$N$ limit of ${\cal N}=4$ $SU(N)$ super Yang-Mills, using the result of the holographic calculation of the interface entropy performed in~\cite{Estes:2014hka}. 

Both for 2$d$ $\mathcal{N}=(2,2)$ and 4$d$ $\mathcal{N}=2$ SCFTs, the K\"ahler potential on the conformal manifold is related to the sphere partition function~$Z[\BS^d]$ as $\log Z[\BS^d] \propto K(\tau,\bar{\tau}) $~\cite{Jockers:2012dk,Gomis:2012wy,Gerchkovitz:2014gta}.
Thus one can relate the interface entropy not just to Calabi's diastasis but also to a ratio of the sphere partition functions in the presence and in the absence of the interface.
Indeed  the paper~\cite{Kobayashi:2018lil} formulated
a relation between the entropy of a conformal defect of general codimension defined in terms of the entanglement entropy and the ratio of the sphere partition functions in the presence and in the absence of the defect.
The main aim of this paper is to derive the formula~(\ref{Conj}), based on a certain assumption,
using CFT techniques similar to~\cite{,,Kobayashi:2018lil}.
We restrict to $\mathcal{N}=2$ superconformal theories realized as gauge theories with Lagrangians, and to marginal couplings identified with complexified gauge couplings, because part of our analysis uses SUSY localization.
It is, however, formally possible to apply the localization to a non-Lagrangian SCFT whose flavor symmetry is gauged by a vector multiplet.
It is conceivable that exactly marginal couplings in $\mathcal{N}=2$ SCFTs can always be realized as gauge couplings.

We summarize the steps for deriving the formula~(\ref{Conj}) as follows.
\begin{enumerate}
\item Based on the replica trick and the Casini--Huerta--Myers map~\cite{Casini:2011kv,Jensen:2013lxa} we show that the interface entropy~(\ref{interface-entropy-def}) is proportional to a ratio of the CFT sphere partition functions in the presence and in the absence of the interface:
\begin{align}\label{eq:relation-ie-partition-function-nonSUSY-intro}
S_{\cal{I}} =\log\left[\frac{ Z^\text{(ICFT)}[\BS^4]}{(Z^\text{(CFT$_+$)}[\BS^4]\,Z^\text{(CFT$_-$)}[\BS^4])^{1/2}}\right] \,.
\end{align}

\item \label{Step:Assumption} We assume that in the presence of a half-BPS superconformal interface $\mathcal{I}$ in an $\mathcal{N}=2$ superconformal field theory, the {\it conformal sphere partition function} defined
 in a conformally invariant scheme
 equals the absolute value of the {\it SUSY sphere partition function} defined
 in a supersymmetric but not necessarily conformally invariant scheme:
\begin{equation}\label{eq:assumption}
\text{ Assumption: } \quad Z^\text{(ICFT)}[\BS^4] = \left|Z^\mathcal{I}_\text{SUSY}[\BS^4] \right| \ .
\end{equation}

\item \label{Step:Localizaton} We show by SUSY localization that the SUSY sphere partition function with a Janus interface is given by the analytic continuation of the sphere partition function without an interface:
\begin{center}
$Z^\mathcal{I}_\text{SUSY}[\BS^4](\tau_+, \bar{\tau}_-)$  is holomorphic in  $\tau_+$ and $\bar{\tau}_-$\ ,
\end{center}
\begin{equation} \label{ICFT-continuation}
Z^\text{(CFT)}[\BS^4](\tau,\bar{\tau}) = Z^\mathcal{I}_\text{SUSY}[\BS^4](\tau_+ = \tau, \bar{\tau}_-=\bar{\tau}) \ .
\end{equation}

\item Use the relation 
\begin{equation}
\log Z^\text{(CFT)}[\BS^4](\tau,\bar{\tau})  =\frac{1}{12}  K(\tau,\bar \tau)
\end{equation}
between the sphere partition function and the K\"ahler potential to derive the relation~(\ref{Conj}).

\end{enumerate}

Our derivation of the relation~(\ref{Conj}) relies on the non-trivial assumption~(\ref{eq:assumption}).
We note, however, that the quantity~(\ref{eq:relation-ie-partition-function-nonSUSY-intro}) with the replacement~(\ref{eq:assumption}) naturally arises if we replace $S^\text{(ICFT)}_E$ by a limit of the supersymmetric R\'enyi entropy, which was introduced in~\cite{Nishioka:2013haa} and is defined using supergravity backgrounds that preserve the supersymmetries used for localization.
Thus even without the assumption~(\ref{eq:assumption}), Calabi's diastasis naturally arises if we use the supersymmetric R\'enyi entropy as an alternative definition of the interface entropy.

In performing SUSY localization, a useful tool is what we call the off-shell construction of supersymmetric defects.
Namely we promote a coupling constant to a supermultiplet (coupling multiplet) and give it a non-trivial spatial profile.
Part of supersymmetry can be preserved by turning on auxiliary fields in the coupling multiplet in such a way that the variations of the fermions vanish.
This method was used in~\cite{Kapustin:2012iw,Okuda:2015yra,Hosomichi:2017dbc,Goto:2018bci,Anderson:2019nlc} for various defects.
Here we apply it to the half-BPS Janus interface in a 4$d$ $\mathcal{N}=2$ gauge theory, which was studied previously based on different constructions~\cite{Gaiotto:2008sd,DHoker:2006qeo,Kim:2008dj,Kim:2009wv,Drukker:2010jp}. 

The outline of this paper is as follows.
In Section~\ref{sec:IE-CFT2} we begin with the discussion of conformal interfaces in general, not necessarily supersymmetric, CFTs.
We define the interface entropy in terms of entanglement entropies and use the Casini--Huerta--Myers map to relate it to a ratio of the sphere partition functions in the presence and in the absence of the interface.
We then explain our assumption~(\ref{eq:assumption}) regarding half-BPS (not necessarily Janus) superconformal interfaces in $\mathcal{N}=2$ SCFTs.
We also explain that this assumption is natural from the point of view of the supersymmetric R\'enyi entropy~\cite{Nishioka:2013haa}.
Section~\ref{sec:off-shell} is devoted to the off-shell construction of the half-BPS Janus interface.
We illustrate the off-shell construction by the simpler case of the flat space, and then construct the Janus interface on $\BS^4$ using off-shell supergravity.
In~Section~\ref{sec:diastasis} we
perform SUSY localization with the Janus interface to show the relation (\ref{ICFT-continuation}).
The relation between the interface entropy and the sphere partition functions is combined with the results of localization to show that the entropy of the Janus interface is proportional to Calabi's diastasis as written in~(\ref{ICFT-continuation}).
In Section~\ref{sec:holography} we perform two holographic computations. 
First, for $\mathcal{N}=4$ $SU(N)$ super Yang-Mills theory, we compute holographically the sphere partition function (or its logarithm, the free energy) in the presence of the Janus interface by evaluating the on-shell action in the supergravity background dual to the interface~\cite{DHoker:2007zhm}.
This involves a certain regularization near the AdS boundary.
Second, again for the $\mathcal{N}=4$ theory, we revisit the computation of the holographic entanglement entropy of the interface, using the same regularization method as for the on-shell action.
The two calculations serve as a check of~(\ref{eq:relation-ie-partition-function-nonSUSY-intro}).
We conclude with discussion in Section~\ref{sec:discussion}.
Appendix~\ref{app:SUSY-SUGRA} collects our conventions and notations, as well as useful facts abound supersymmetry and supergravity.
Appendices~\ref{app:conf} and~\ref{sec:SRE-details} contain technical details that we use in the main text.

\section{Interface entropies in CFT and SCFT}\label{sec:IE-CFT2}


In this section we define the interface entropy in terms of entanglement entropies and relate it to a ratio of the sphere partition functions in the presence and in the absence of the interface.
We also explain our assumption~(\ref{eq:assumption}) regarding half-BPS superconformal interfaces in $\mathcal{N}=2$ SCFTs.

\subsection{Entanglement entropy in the presence of an interface}\label{sec:EE-int}

We begin by reviewing the standard definition of the entanglement entropy, with a conformal interface included in a straightforward way.
For a similar discussion with defects of general codimensions, see~\cite{Kobayashi:2018lil}. 

We consider a 4$d$ CFT in Minkowski space with coordinates $(t,y^1,y^2,y^3)$.
Let us introduce along the hyperplane~$y^3=0$  a \emph{conformal interface} $\CI$ that preserves a subgroup~$SO(2, 3)$ of the conformal group~$SO(2,4)$.
  We also use spherical coordinates $(r,\phi,\chi)$ related to the Cartesian coordinates as $(y^1,y^2,y^3)=(r\sin\phi\cos\chi, r\sin\phi\sin\chi,r\cos\phi)$.
Let us take the entangling surface $\Sigma$ to be a 2-sphere with radius~$R$ inside the $t=0$ time slice
\begin{align}
	\Sigma = \{ t=0,\, r= R\} \ .
\end{align}
We decompose the Hilbert space modified by ${\mathcal I}$,~$\mathcal{H}_{\mathcal{I}}$, into the tensor product of~$\mathcal{H}_A$ and~$\mathcal{H}_B$ that correspond to the regions $r<R$ and $r>R$ in the constant time slice $\mathbb{R}^3$ at $t=0$, respectively:%
\footnote{%
We choose not to delve into to the subtleties associated with such a decomposition for a gauge theory.
}
\begin{equation}
\mathcal{H}_{\mathcal{I}} = \mathcal{H}_A \otimes \mathcal{H}_B \,.
\end{equation}
Inside the~$t=0$ slice, the entangling surface $r=R$ intersects the interface along the great circle at $\phi = \pi/2 $.
Using the ground state $|0\rangle \in \mathcal{H}_{\mathcal{I}} $ we form the density matrix $\rho={\rm tr}_{B} |0\rangle\langle0| $ by the partial trace over $ \mathcal{H}_B$.
Next, by taking the partial trace over $\mathcal{H}_A$ we define the entanglement entropy
\begin{equation}
S_E^\text{(ICFT)} := - {\rm tr}_{A}\, \rho \log \rho\ ,
\end{equation}
and the R\'eny entropy
\begin{equation}\label{Renyi_definition}
S_n^\text{(ICFT)} := \frac{1}{1-n} \log {\rm tr}_{A}\, \rho^n \,.
\end{equation}
The two quantities are related as
\begin{equation}\label{ICFT_RE-limit}
\lim_{n\rightarrow 1}S_n^\text{(ICFT)}  =S_E^\text{(ICFT)} \ .
\end{equation}
By construction $S_E^\text{(ICFT)} $ and $S_n^\text{(ICFT)}$ are non-negative.

The replica trick identifies the quantity ${\rm tr}_{A}\, \rho^n$ with the partition function $Z[\CM_n]$, {\it i.e.}, the  path integral on the $n$-fold branched cover $\mathcal{M}_n$ of the Euclidean space $\mathbb{R}^4$, normalized by~$Z[\CM_1]^n$:
\begin{equation} \label{rho-Z-ratio}
{\rm tr}_{A}\, \rho^n = \frac{Z[\CM_n]}{Z[\CM_1]^n} \ .
\end{equation}
Since we are interested in the continuous limit $n\rightarrow 1$, we wish to define $\mathcal{M}_n$ for non-integer~$n$.

\begin{figure}[tb]
\begin{center}
\vspace{-1.5cm}
  \includegraphics[width=10cm]{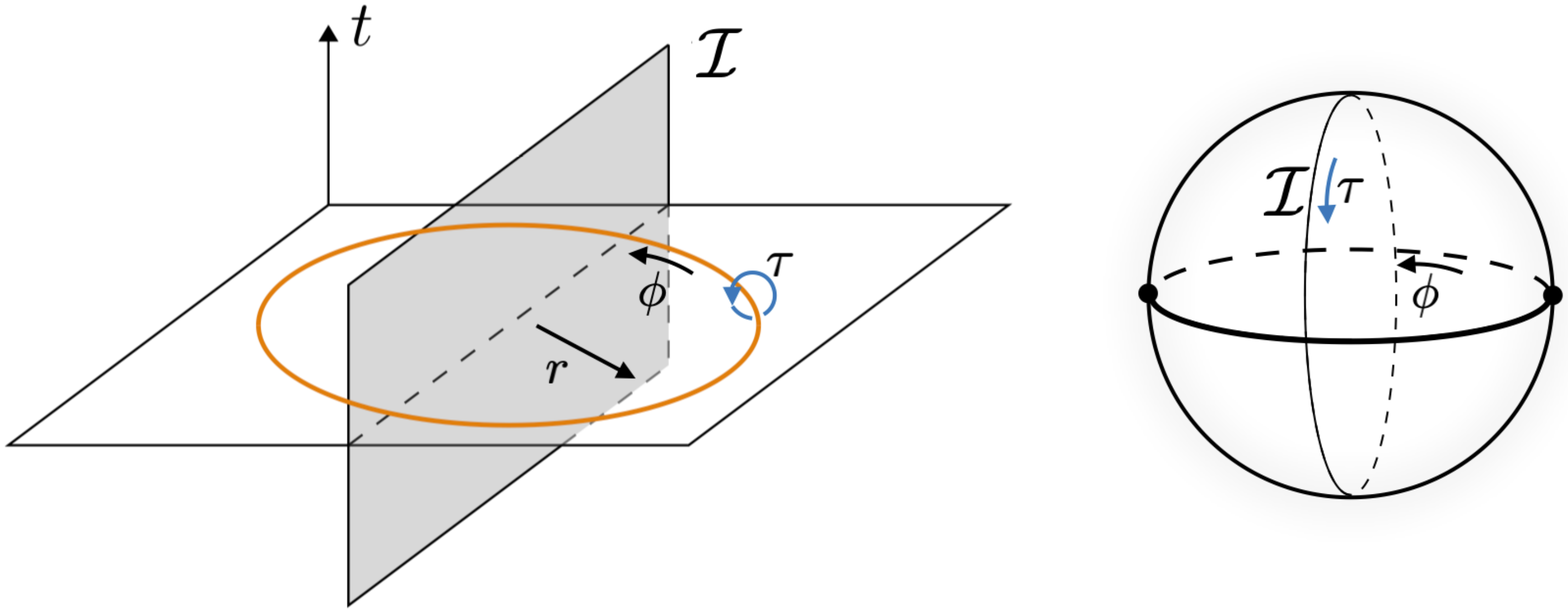}
  \vspace{-1.5cm}
  \caption{(Left) A codimension-one conformal interface ${\cal I}$ and the entangling surface (within the constant time slice $t=0$) in the 4$d$ Euclidean spacetime $\BR^{4}$.
  The $\tau$ direction (blue arrow) corresponds to the modular flow.  (Right) The conformal interface extends along the equator $\BS^3$ at $\phi=\pi/2$ of the 4-sphere $\BS^4$. 
   }
   \label{fig:CHM}
 \end{center}
\end{figure}

A useful tool to achieve this is the so-called Casini--Huerta--Myers map~\cite{Casini:2011kv,Jensen:2013lxa}.
We perform the Wick rotation via the substitution $t\rightarrow -  {\rm i} t$ and consider the Euclidean space with coordinates $(t,y^1,y^2,y^3)$ and the metric
\begin{align}
\begin{aligned}
	\d s_{\BR^{4}}^2 
			&= \d t^2 + \d  r^2 +  r^2\,\left(\d \phi^2 + \sin^2\phi\,\d \chi^2 \right)   \ .
\end{aligned}
\end{align}
Let us perform a change of coordinates to $(\tau, \theta,\phi,\chi)$ via (the Euclidean version of) the Casini--Huerta--Myers (CHM) map%
\footnote{%
We believe that the reader can distinguish, based the context, the coordinate~$\tau$ from the coupling~$\tau$. 
}
\begin{align}\label{CHM-Euc}
	\begin{aligned} 
	t &=R\, \frac{\sin\theta\sin \tau}{1+\sin\theta\cos \tau}\ , \\ 
	 r &=R\, \frac{\cos\theta}{1+\sin\theta\cos \tau } \ .
	\end{aligned}
\end{align}
Through this, the Euclidean space is conformally equivalent to the round sphere as
\begin{align}
	\d s_{\BR^{4}}^2 
	=\Omega^2\d s_{\BS^4}^2
	  \ ,
\end{align}
with the conformal factor 
\begin{align}
	\Omega = \frac{R}{1+\sin\theta\cos\tau}	
\end{align}
and the round sphere metric
\begin{align} \label{eq:S4-metric}
	\d s_{\BS^4}^2
	=\d \theta^2+\sin^2\theta\, \d \tau^2  + \cos^2\theta\,\left(\d \phi^2 + \sin^2\phi\,\d \chi^2
	 \right)\ .
\end{align}
The entangling surface~$\Sigma$ is mapped to the 2-sphere at $\theta=0$.
The translation in the $\tau$ direction fixes $\Sigma$ and corresponds to the modular flow generated by the modular Hamiltonian~$H$ defined by $\rho=e^{-H}$~\cite{Casini:2011kv}.
See Figure~\ref{fig:CHM}.
The $n$-fold cover $\mathcal{M}_n$ has the metric
\begin{equation}\label{CHM_branched}
	\d s_{\mathcal{M}_n}^2 
	=\Omega^2\d s_{\BS^4_n}^2 \ ,
\end{equation}
with
\begin{align}\label{Branched_Sphere}
		\d s_{\BS^4_n}^2 &= \d \theta^2 + n^2\sin^2\theta\, \d \tau^2 + \cos^2\theta\,\left(\d \phi^2 + \sin^2\phi\,\d \chi^2 \right) \ .
\end{align}
The range of $\tau$ is $0\le \tau < 2\pi$.
This metric is singular for $n\neq 1$.

\subsection{Interface entropy and the sphere partition function}\label{sec:IE-sphere}

Armed with the CHM map \eqref{CHM_branched} associating the replica space $\CM_n$ to the $n$-fold cover of a 4-sphere $\BS_n^4$, we will derive a relation between the entanglement entropy and the sphere partition function in ICFT.
While we are only concerned with ICFT in four dimensions there is no difficulty in repeating the same argument for the CHM map in general $d$ dimensions (just by replacing the entangling region $\BS^2$ with $\BS^{d-2}$).
So we closely follow the derivation in \cite{Kobayashi:2018lil} which uses the dimensional regularization for calculating the entanglement entropy in CFT with conformal defects for a moment.
This approach is not only general enough, but also simplifies the derivation by avoiding an extra care for conformal anomalies as they are automatically incorporated into poles at even dimensions.
We defer the discussion about conformal anomalies in ICFT to Section \ref{sec:anomaly-interface}.

In the dimensional regularization we adopt a scheme such that the theory is strictly conformal even at quantum level.
In other words, we start with an odd-dimensional CFT without conformal anomalies and analytically continue it to general dimensions.
Hence the CFT partition functions, even in the presence of an interface, on the $n$-fold covers of the Euclidean space and $d$-sphere are the same under conformal transformation of the type \eqref{CHM_branched}:
\begin{align} 
	Z^\text{(ICFT)}[\CM_n] = Z^\text{(ICFT)}[\BS^d_n]
		 \ .
\end{align}
Note that the equality between the two partition functions holds only up to power-law UV divergences.
It follows from this relation together with \eqref{Renyi_definition} and \eqref{rho-Z-ratio} that the R{\'e}nyi entropy across a sphere in ICFT is given by
\begin{align}\label{SphereRE}
	S_n^\text{(ICFT)} = \frac{1}{1-n}\, \log \frac{Z^\text{(ICFT)}[\BS^d_n]}{\left( Z^\text{(ICFT)}[\BS^d]\right)^n} \ .
\end{align}
We note that this expression is trivially valid in the absence of an interface.

Now we consider an interface CFT built out of two CFTs, CFT$_+$ and CFT$_-$, glued together along the interface $\CI$.
We define the \emph{interface entropy} as the contribution to the entanglement entropy by the interface $\CI$:
\begin{align}\label{Def_IE}
	S_\CI \equiv S_{E}^\text{(ICFT)} - \frac{S_{E}^\text{(CFT$_+$)} + S_{E}^\text{(CFT$_-$)}}{2}\ .
\end{align}

Using~(\ref{ICFT_RE-limit}) and~(\ref{SphereRE}) we can write the quantity $S_{E}^\text{(ICFT)}$ in~(\ref{Def_IE}) as
\begin{equation}\label{ICFT_RE-derivative}
S_{E}^\text{(ICFT)} = 
\log\, Z^\text{(ICFT)}[\BS^d] -\left. \partial_n \log  Z^\text{(ICFT)}[\BS^d_n] \right|_{n=1}\,.
\end{equation}
We wish to show that the second term in~(\ref{ICFT_RE-derivative}) vanishes, {\it i.e.}, that the relation
\begin{equation}\label{ICFT_EE-sphere}
S_{E}^\text{(ICFT)} = 
\log\, Z^\text{(ICFT)}[\BS^d] 
\end{equation}
holds.
For this we need the behavior of the R\'enyi entropy \eqref{SphereRE} in ICFT at $n=1+\epsilon$ with small $\epsilon$.  
In the framework of general, not necessarily supersymmetric, conformal field theory, 
$\log Z^\text{(ICFT)}[\BS^d_n]$ and  $\log Z^\text{(ICFT)}[\BS^d]$ differ by the variation of the background metric $\delta g_{\tau\tau} = (n^2 -1)\, \sin^2\theta$.
In terms of the stress tensor 
defined by
\begin{align}
\langle	T^{\mu\nu}  \rangle = -\frac{2}{\sqrt{g}}\, \frac{\delta\, \log Z} {\delta g_{\mu\nu}}\ ,
\end{align}
where $Z$ is a general partition function that depends on the metric,
we can write
\begin{equation}\label{LogZ_expansion-nonSUSY}
	-\log\, Z^\text{(ICFT)}[\BS^d_n] +\log\, Z^\text{(ICFT)}[\BS^d]
	= +\frac{1}{2}\int_{\BS^d}\,  \delta g_{\mu\nu}\, \langle\, T^{\mu\nu}\,\rangle_{\BS^d}^\text{(ICFT)}
+{\cal O}(\epsilon^2) \ .
\end{equation}
To study the one-point function of the stress tensor we use a conformal mapping between~$\BS^d$ and the flat space.
This map may be but does not have to be the CHM map~(\ref{CHM-Euc}).
In general the one-point function~$ \langle\, T^{\mu\nu}\,\rangle_{\BS^d}^\text{(ICFT)}$ transforms under a conformal transformation as
\begin{align}\label{VEV_stress_tensor}
	\langle\, T_{\mu\nu}\,\rangle^{(\text{ICFT})}_{\BS^d} =(\text{Weyl factor})^2 \langle\, T_{\mu\nu}\,\rangle^{(\text{ICFT})}_{\BR^d} \ .
\end{align}
One can easily show $\langle\, T_{\mu\nu}\,\rangle^{(\text{ICFT})}_{\BR^d}$ vanishes due to the residual conformal symmetry $SO(1,4)$ preserved by the interface \cite{McAvity:1995zd,Billo:2016cpy}, so we conclude that
the interface entropy is given by the  combination
\begin{align}\label{eq:relation-ie-partition-function-nonSUSY}
S_{\cal{I}} =\log\left[\frac{ Z^\text{(ICFT)}[\BS^d]}{(Z^\text{(CFT$_+$)}[\BS^d]\,Z^\text{(CFT$_-$)}[\BS^d])^{1/2}}\right] \ ,
\end{align}
of the sphere partition functions with and without an interface.
In what follows we will use this relation in the calculation of the interface entropy in $d=4$ dimensions.

\subsection{Interface entropy in SCFT}\label{sec:int-ent-SCFT}

We now turn to half-BPS superconformal interfaces in 4$d$ $\mathcal{N}=2$ superconformal field theories.
For our conventions, see Appendix~\ref{app:notations}.

In flat space with Cartesian coordinates~$y^\mu$ the Poincar\'e supersymmetry and special superconformal transformations are parametrized as $\delta_Q=\bar{\epsilon}{}^i Q_i + \bar{\epsilon}_i Q^i$ and $\delta_S=\bar{\eta}{}^i S_i + \bar{\eta}_i S^i$, where a bar on a 4-component spinor parameter indicates the Weyl conjugate defined in~(\ref{Weyl-conj-def-Psi}).%
\footnote{%
The parameters here are related to the parameters in Appendix~\ref{sec:N=2mult} as $(\epsilon^i,\epsilon_i)_\text{there} = (\epsilon^i + y^\mu\gamma_\mu \eta^i, \epsilon_i + y^\mu \gamma_\mu \eta_i)$.
}
The spinors $\epsilon^i$ and $\eta_i$ are left-handed, while~$\epsilon_i$ and $\eta^i$ are right-handed.
The operators $Q_i$ and $S^i$ are left-handed, while~$Q^i$ and $S_i$ are right-handed.
A half-BPS superconformal interface at $y^3=0$ preserves the fermionic symmetries with parameters satisfying
\begin{equation} \label{half-BPS-epsilon-eta}
\epsilon_i = \rho_{ij}\gamma^3 \epsilon^j  \ , \quad
\eta_i = - \rho_{ij}\gamma^3 \eta^j  \ ,
\end{equation}
where the fixed symmetric tensor $\rho_{ij}$ satisfies $\rho_{ij} \bar \rho^{jk}=\delta_i^k$ with $\bar\rho^{ij}:=(\rho_{ij})^*$.%
\footnote{%
Such $\rho_{ij}$ can be parametrized as $\rho_{ij} = e^{{\rm i}\alpha} \vec n\cdot\vec\tau_{ij}$, where $\alpha$ is real and $\vec n$ is a real unit vector.
They transform under $U(1)_R$ and $SU(2)_R$.
}
In other words, the preserved supercharges and special superconformal charges are
\begin{equation}
Q_i - \rho_{ij} \gamma_3 Q^j \ , \quad S_i +\rho_{ij}\gamma_3 S^j \, .
\end{equation}
They generate the 3$d$ $\mathcal{N}=2$ superconformal algebra $OSp(2|4)_{sc}$.

Since such an interface is a special kind of conformal interface, our discussion in Sections~\ref{sec:EE-int} and~\ref{sec:IE-sphere} applies to it.
There are, however, two important differences between the conformal case and the superconformal case.

The first difference is that superconformal field theories and interfaces naturally couple to background supergravity (or conformal supergravity) fields other than the metric.
The partition functions are functionals of these fields.
In general a supersymmetric background involves non-zero supergravity fields.%
\footnote{%
In the supersymmetric $\BS^4$ background, the metric is the only non-zero field in the Poincar\'e supergravity multiplet~\cite{Hama:2012bg,Pestun:2014mja}.  There are non-zero fields in compensating multiplets~\cite{Gomis:2014woa} that violate conformal invariance and unitarity.
See~(\ref{Vc1}) and (\ref{Vc2}).
}

The second difference is that the counterterms dictated by supersymmetry involve supergravity fields other than the metric.
When we turn off supergravity fields other than the metric, as in the supersymmetric $\BS^4$ background, such terms reduce to non-SUSY counterterms that involve only the metric (and other non-supergravity background fields), but their coefficients are related by supersymmetry.
This mechanism gives universal meanings to some, a priori non-universal, terms in the effective action~\cite{Gerchkovitz:2014gta}.

To establish the relation~(\ref{Conj}) between the interface entropy and Calabi's diastasis, an important step for us---Step~\ref{Step:Localizaton} in the introduction---involves localization that computes the supersymmetric partition function $Z^\mathcal{I}_\text{SUSY}[\BS^4]$ of the system with an interface in a supersymmetric background.
As we will see in Section~\ref{sec:diastasis}, the SUSY partition function~$Z^\mathcal{I}_\text{SUSY}[\BS^4]$  is in general complex.
On the other hand, so far we have related the interface entropy only to the conformal partition function $Z^\text{(ICFT)}[\BS^4]$, which is real and positive by unitarity.

Based on these motivations we make the assumption~(\ref{eq:assumption}) in Step~\ref{Step:Assumption}, {\it i.e.},
\begin{equation}\label{eq:assumption-Sec2}
\quad Z^\text{(ICFT)}[\BS^4] = \left|Z^\mathcal{I}_\text{SUSY}[\BS^4] \right|.
\end{equation}
Combined with~(\ref{eq:relation-ie-partition-function-nonSUSY}), this gives the interface entropy
\begin{align}\label{eq:relation-ie-partition-function}
	S_{\cal{I}} =\log\left[\frac{\left| Z^\mathcal{I}_\text{SUSY}[\BS^4] \right|}{(Z^\text{(CFT$_+$)}_\text{SUSY}[\BS^4]\,Z^\text{(CFT$_-$)}_\text{SUSY}[\BS^4])^{1/2}}\right]\ ,
\end{align}
in terms of supersymmetric sphere partition functions with and without an interface.
We note that the combination~(\ref{eq:relation-ie-partition-function}) coincides with the ``boundary free energy'' considered in~\cite{Gaiotto:2014gha,DiPietro:2019hqe,Gupta:2019qlg}.%
\footnote{%
In 3$d$ it is common to define the free energy as $F=-\log |Z_\text{SUSY}[{\BS^3}]|$ in terms of the absolute value of the partition function computed by SUSY localization.  See for example~\cite{Jafferis:2011zi}.}

We explain in Section~\ref{sec:super-Weyl} that one can use the super-Weyl anomaly of~\cite{Bachas:2016bzn} to prove the 2$d$ version of the assumption~(\ref{eq:assumption-Sec2}).

\subsection{Interface entropy and the supersymmetric R\'enyi entropy}\label{sec:SRE}

We now explain that the assumption~(\ref{eq:assumption-Sec2}) is natural from the point of view of the supersymmetric R\'enyi entropy~\cite{Nishioka:2013haa}.
More precisely~(\ref{eq:assumption-Sec2}) is equivalent to the statement that the entanglement entropy~$S_E^\text{(ICFT)}$ coincides with the $n\rightarrow 1$ limit of the supersymmetric R\'enyi entropy~$S_{{\rm SUSY}\ n}^\text{(ICFT)}$ that we define below.

Even in the presence of a conformal interface, one can relate the (ordinary) R\'enyi entropy to the partition function on the $n$-fold covering of the round sphere, as we wrote in~(\ref{SphereRE}).
This expression is somewhat formal because we do not specify how we deal with the conical singularities for $n\neq 1$.
One can make it more precise by considering a supersymmetric background~$\widetilde{\BS}^4_n$ that regularizes the $n$-fold covering~$\BS^4_n$~\cite{Huang:2014pda,Pestun:2014mja}.
We review the supergravity background~$\widetilde{\BS}^4_n$ in Appendix~\ref{sec:SRE-background}.%
\footnote{%
Although we do not show this explicitly, we expect that 
in the supersymmetric~$\widetilde{\BS}^4_n$ background one can construct a SUSY preserving Janus interface that reduces to the half-BPS interface in the $n\rightarrow 1$ limit.
The worldvolume of the interface is invariant under the Killing vector generated by the square of the supercharge preserved by the background.
The $\widetilde{\BS}^4_n$ background is a member of the more general family of supersymmetric backgrounds that includes the ellipsoid of~\cite{Hama:2012bg}, for which a Janus interface has a natural interpretation in the context of the AGT correspondence~\cite{Drukker:2010jp}.
}
In the limit $n\rightarrow 1$ the background reduces to the round sphere $\BS^4$ with all supergravity fields other than the metric vanishing.
Let us denote the partition function for $\widetilde{\BS}^4_n$ by~$Z^\mathcal{I}_\text{SUSY}[\widetilde\BS^4_n]$ and define the supersymmetric R\'enyi entropy
\begin{align}\label{SRE-absolute-value}
S_{{\rm SUSY}\ n}^\mathcal{I} := \frac{1}{1-n}\, {\rm Re} \log \frac{Z^\mathcal{I}_\text{SUSY}[\widetilde\BS^4_n]}{\big( Z^\mathcal{I}_\text{SUSY}[\BS^4]\big)^n}\ .
\end{align}
We take the real part of the logarithm, or equivalently the absolute value inside the logarithm, mimicking the original definition in 3$d$ (without an interface)~\cite{Nishioka:2013haa}.
(See also \cite{Huang:2014gca,Nishioka:2014mwa}).
The supersymmetric R\'enyi entropy is a natural and meaningful physical quantity in general dimensions~\cite{Huang:2014pda,Crossley:2014oea,Hama:2014iea,Alday:2014fsa,Giveon:2015cgs,Mori:2015bro,Zhou:2015kaj,Nian:2015xky,Nishioka:2016guu,Yankielowicz:2017xkf}.

If we assume that the entanglement entropy is related to the supersymmetric R\'enyi entropy as
\begin{equation} \label{eq:EE-SRE-limit}
S_{E}^\text{(ICFT)}  = \lim_{n\rightarrow 1}S_{{\rm SUSY}\ n}^\mathcal{I} \ , 
\end{equation}
we have the supersymmetric version of the equality~(\ref{ICFT_RE-derivative}):
\begin{equation}\label{ICFT_SRE-derivative}
S_{E}^\text{(ICFT)} = 
\log \left| Z^\mathcal{I}_\text{SUSY}[\BS^4] \right| - \partial_n \, {\rm Re}\, \log  \left. Z^\mathcal{I}_\text{SUSY}[\widetilde \BS^4_n] \right|_{n=1}\ .
\end{equation}
In Appendix~\ref{ss:supercurrent} we show that the second term vanishes.
Thus
\begin{equation}\label{ICFT_SRE}
S_{E}^\text{(ICFT)} = 
\log \left| Z^\mathcal{I}_\text{SUSY}[\BS^4] \right|  \ .
\end{equation}
Comparing~(\ref{ICFT_SRE}) with~(\ref{ICFT_EE-sphere}), we see that~(\ref{eq:EE-SRE-limit}) is equivalent to the assumption~(\ref{eq:assumption-Sec2}).


\section{Off-shell construction of the Janus interface}\label{sec:off-shell}

In this section we provide an off-shell construction of the Janus interface in a general $\mathcal{N}=2$ SCFT in flat space and on $\mathbb{S}^{4}$.
We borrow tools from $\mathcal{N}=2$ supergravity.
Supersymmetry transformations of the relevant supermultiplets are summarized in Appendix~\ref{sec:N=2mult}.

\subsection{Off-shell construction in flat space}\label{sec:off-shell-flat}

Let us illustrate the off-shell construction method of the Janus interface in a general $\mathcal{N}=2$ SCFT by first considering the simpler set-up of Minkowski space with coordinates~$y^\mu$.
While the physical reality conditions are clearer in Minkowski signature~(see~\cite{Freedman:2012zz}), all the formulas in this subsection are also valid in Euclidean signature.
Without loss of generality we focus on a single marginal coupling~$\tau$.

A crucial ingredient is the coupling chiral multiplet of Weyl weight zero
\begin{equation} \label{coupling-m-chiral}
\mathcal{T}=(\tau,\Psi^{(\tau)}_{ i},B_{ij}^{(\tau)},F_{\mu\nu}^{(\tau)-},\Lambda_{ i}^{(\tau)}, C^{(\tau)})  \ .
\end{equation}
It is accompanied by an anti-chiral multiplet
\begin{equation}\label{coupling-m-anti-chiral}
\overline{\mathcal{T}}=(\overline{\tau},\,{\Psi^{(\tau)i}},\,B^{(\tau)ij},\,F_{\mu\nu}^{(\tau)+},\,{\Lambda^{(\tau)i}},\,\overline{C}{}^{(\tau)}) \ ,
\end{equation}
where we take $\bar\tau$ to be the complex conjugate of $\tau$: $\bar\tau=\tau^*$.
See Appendix~\ref{app:SUSY-SUGRA} for our conventions.
We wish to construct an interface characterized by a general profile of the complexified coupling $\tau(y^3)$ with part of Lorentz symmetry unbroken.
We set the fermions in the coupling multiplet to zero.
To preserve some supersymmetry, we require the auxiliary fields in $\mathcal{T}$ to take appropriate values so that the variations of the fermions vanish.
Using the unbroken Lorentz symmetry we obtain, for constant $\epsilon^i$ and $\epsilon_i$,
\begin{align}
\delta\Psi_{i}^{(\tau)}&=
(\partial_3 \tau)  \gamma^3   \epsilon_{i}+\frac{1}{2}\,B^{(\tau)}_{ij}\epsilon^{j} \ ,
\\
\delta \Lambda_{i}^{(\tau)}&= -\frac12 \partial_3 B_{ij}^{(\tau)}  \varepsilon^{jk} \gamma^3 \epsilon_k + \frac12 C^{(\tau)} \varepsilon_{ij} \epsilon^j \ ,
\\
\delta\Psi^{(\tau)i}&=
(\partial_3 \overline\tau)\gamma^3   \epsilon^{i}+\frac{1}{2}\,B^{(\tau)ij}\epsilon_{j} \ ,
\\
\delta \Lambda^{(\tau)i}&= -\frac12 \partial_3 B^{(\tau)ij}  \varepsilon_{jk} \gamma^3 \epsilon^k + \frac12 \overline{C}{}^{(\tau)} \varepsilon^{ij} \epsilon_j \ .
\end{align}
We demand that these expressions vanish on a half-dimensional subspace of the space of $(\epsilon^i,\epsilon_i)$.
As functions of $y^3$, $B^{(\tau)}_{ij}$ must be proportional to $\partial_3 \tau$,  $C^{(\tau)}$ to $\partial_3^2 \tau$, $B^{(\tau)ij}$ to $\partial_3 \overline\tau$, and $\overline C{}^{(\tau)}$ to $\partial_3^2 \overline\tau$.
The solutions are parametrized by a $U(1)$ phase $e^{\i\alpha}$ and a real unit vector $\vec n$, which naturally transform under $U(1)_R$ and $SU(2)_R$, respectively.
We write
\begin{equation} \label{rho-alpha-n}
 \rho_{ij} = e^{{\rm i}\alpha}\,\vec n \cdot\vec\tau_{ij} \ , \qquad \overline\rho^{ij} = e^{-{\rm i}\alpha}\,\vec n\cdot \vec\tau^{\, ij} \ .
\end{equation}
Then
\begin{align}
\epsilon_i &= \rho_{ij} \gamma^3 \epsilon^j \ , &  &  \label{epsilon-Janus-flat}\\
B^{(\tau)}_{ij} &= -2  \rho_{ij} \,\partial_3\tau \ , &
B^{(\tau)ij} &= -2 \overline\rho{}^{ij}\, \partial_3\overline\tau \ , \label{B-flat}\\
C^{(\tau)}  &= -2 e^{+2{\rm i}\alpha}\, \partial_3^2\tau \ , &
\overline{C}{}^{(\tau)}  &= -2 e^{-2{\rm i}\alpha}\, \partial_3^2\overline \tau \  . \label{C-flat}
\end{align}
We note that~(\ref{epsilon-Janus-flat}) coincides with the first equation in~(\ref{half-BPS-epsilon-eta}).

We now specialize to a step function profile
\begin{align}\label{tau-flat-step}
\tau(y^3)
=
\begin{cases}
\tau_{+} & \text{for} \quad y^{3}>0\ ,
\\
\tau_{-} & \text{for} \quad y^{3}<0\ .
\end{cases}
\end{align}
Let us define $\Delta\tau:=\tau_+-\tau_-$.
In the expressions for the auxiliary fields in~(\ref{B-flat}) and~(\ref{C-flat}), we get~$\partial_3\tau= \Delta\tau\,\delta(y^3)$, $\partial_3^2\tau= \Delta\tau\,\delta'(y^3)$, where the prime denotes the derivative.
Explicitly,
\begin{equation}
\label{BC-flat-delta}
\begin{aligned}
B^{(\tau)}_{ij} &= -2  \rho_{ij} \Delta\tau\, \delta(y^3) \ , &\quad
B^{(\tau)ij} &= -2 \overline\rho{}^{ij}  \Delta\overline{\tau}\, \delta(y^3) \ ,\\
C^{(\tau)}  &= -2 e^{+2{\rm i}\alpha} \Delta\tau  \,\delta'(y^3) \ , &\quad
\overline{C}{}^{(\tau)}  &= -2 e^{-2{\rm i}\alpha} \Delta\overline{\tau} \, \delta'(y^3)\  . 
\end{aligned}
\end{equation}
We are interested in special superconformal transformations, which we denote by~$\delta_\eta$.
We take~$\eta^i$ and~$\eta_i$ constant and make substitutions $\epsilon^i\rightarrow y^\mu\gamma_\mu \eta^i$ and $\epsilon_ i\rightarrow y^\mu\gamma_\mu \eta_i$ in~(\ref{eq:susy-variation-psi}) and~(\ref{eq:susy-variation-lambda})  to get $\delta_\eta \Psi_{i}^{(\tau)}= 0$ and
\begin{align}
\delta_\eta \Lambda_{i}^{(\tau)}&=  -\frac12\, \partial_3 B^{(\tau)}_{ij} \varepsilon^{jk} y^\mu \gamma^3\gamma_\mu\eta_k + \frac12\, C^{(\tau)} \varepsilon_{ij} y^\mu\gamma_\mu\eta^j - B^{(\tau)}_{ij} \varepsilon^{jk} \eta_k \nonumber
\\
&=2 \Delta\tau\,
\partial_3(y^3\delta(y^3)) \rho_{ij}\, \varepsilon^{jk}\eta_k \label{delta-eta-Lambda}
\\
&\qquad
  -  \Delta\tau\, e^{2{\rm i}\alpha}\, \delta'(y^3)  \Big( y^3+\sum_{a=0}^2 y^a  \gamma_a\gamma^3\Big)  \varepsilon_{ij} \overline{\rho}{}^{jk} ( 
  \eta_k +\rho_{kl} \gamma^3\eta^l)
\ .\nonumber
\end{align}
As a distribution, {\it i.e.}, as a linear functional on the space of smooth functions with compact support, $\partial_3(y^3\delta(y^3))$ is zero.
Then~$\delta_\eta \Lambda_{i}^{(\tau)}$ vanishes precisely when the second equation in~(\ref{half-BPS-epsilon-eta}) is satisfied.
The same is true for~$\delta_\eta \Psi^{(\tau)i}$ and $\delta_\eta \Lambda^{(\tau)i}$, which are obtained from~(\ref{delta-eta-Lambda}) by charge conjugation.

Thus we succeeded in constructing a half-BPS superconformal Janus interface in flat Minkowski space by an off-shell method.
It preserves the subalgebra $OSp(2|4)_{sc}$ of the 4$d$ $\mathcal{N}=2$ superconformal algebra~$SU(2,2|2)$.%
\footnote{%
Our notations do not distinguish different real forms of the algebras that arise in Minkowski and Euclidean signatures.
We also use group (capital letter) notations even though we really mean Lie algebras.
}
The former is the 3$d$ $\mathcal{N}=2$ superconformal algebra.
We note that the background values of the coupling multiplet in flat space respect the physical reality conditions, {\it i.e.}, $B^{(\tau)ij}=(B^{(\tau)}_{ij})^*$, $(C^{(\tau)})^*=\overline{C}{}^{(\tau)}$.

\subsection{Massive superalgebra on~$\mathbb{S}^{4}$}\label{sec:massive}

Because~$\mathbb{S}^{4}$ is conformally flat, the full $\mathcal{N}=2$ superconformal algebra on $\mathbb{S}^{4}$ is again $SU(2,2|2)$.
Similarly a half-BPS superconformal interface along~$\mathbb S^3\subset\mathbb S^4$ preserves the 3$d$ $\mathcal{N}=2$ superconformal algebra~$OSp(2|4)_{sc}$.
Another relevant algebra is the massive superalgebra~$OSp(2|4)_{m}$ generated by the SUSY parameters~\cite{Gomis:2014woa}
\begin{align} \label{epsilon-massive}
\e^{i}= 
{ e^{-\frac{\rm i}{2} \beta}}
P_L \chi^{i}\ ,
\qquad
\e_{i}=
{ e^{\frac{\rm i}{2} \beta}}\,
\vec{n}\cdot\vec{\tau}_{ij} \, P_R \chi^{j}\ ,
\end{align}
where
$\chi^{i}$ is a Killing spinor satisfying
\begin{align}\label{chi-Killing}
\nabla_{\mu}\chi^{i}
=
\frac{\mathrm{i}}{2r}\gamma_{\mu}\chi^{i}\ .
\end{align}
Here $r$ is the radius of $\BS^4$ and $\vec{n}$ is a unit three-vector, which we will identify with the vector denoted by the same symbol in~(\ref{rho-alpha-n}) when we introduce a Janus interface.
We also introduced a $U(1)_R$ phase $\beta$.

We take the stereographic coordinates $x^\mu$ and set
$x:=(\sum (x^\mu)^2)^{1/2} $.
The metric is given by
\begin{align}\label{round-metric}
g_{\mu\nu}=f(x)^{2}\,\delta_{\mu\nu}\ ,
\qquad
f(x)=\frac{1}{1+\frac{x^{2}}{4r^{2}}}\ .
\end{align}
The gamma matrices in upper and lower cases are related by the vielbein as $\gamma^\mu =  \Gamma^a e_a{}^\mu$, with~$\Gamma^a$ being constant gamma matrices satisfying~$\Gamma^a\Gamma^b + \Gamma^b\Gamma^a = 2\delta^{ab}$, and the vielbein given by~$e_a{}^\mu= f(x)\delta_a^\mu$.
In the stereographic coordinates~$x^\mu$, the Killing spinors can be written as
\begin{align}\label{eq:Killing-spinor-stereographic}
\chi^{j}=\sqrt{f}\left(1+\frac{\mathrm{i}}{2r}x_{\mu}\Gamma^{\mu}\right)\chi_{0}^{j}\ ,
\end{align}
where $\chi_{0}^{j}$ is a constant spinor.
Then we can write $\e^{i},\e_{i}$ as 
\begin{align} \label{epsilon-chi}
\e^{i}
&={ e^{-\frac{\rm i}{2} \beta}}
\sqrt{f}\left( P_L \chi_{0}^{i}+\frac{\mathrm{i}}{2r}x_{\mu}\Gamma^{\mu} P_R \chi_{0}^{i}\right)\ ,
\\
\e_{i}
&=
{ e^{\frac{\rm i}{2} \beta}}
\sqrt{f} \, \vec{n}\cdot\vec{\tau}_{ij}\left( P_R \chi_{0}^{j}+\frac{\mathrm{i}}{2r}x_{\mu}\Gamma^{\mu} P_L \chi_{0}^{j}\right)\ .
\end{align}

If we further restrict the symmetry by imposing the chirality condition
\begin{align} \label{eq:chi-zero-chirality}
P_{L}\chi_{0}^i=0\ ,
\end{align}
then the corresponding symmetry is $OSp(2|2)_m$~\cite{Gomis:2014woa}.
We do not lose generality by imposing this condition, as we will explain in Section~\ref{sec:dependence}.
It will, however, also be useful to consider an alternative choice of massive subalgebra given by replacing~(\ref{eq:chi-zero-chirality}) with
\begin{align} \label{eq:chi-zero-chirality-alternative}
\text{(alternative)}\qquad P_{R}\chi_{0}^i=0\ .
\end{align}

\subsection{Off-shell construction on $\mathbb{S}^{4}$}\label{subsec:off-shell-construction-sphere}

We now perform the off-shell construction of the 
Janus interface on $\BS^4$.
As in Section~\ref{sec:off-shell-flat} this is done by introducing the coupling chiral multiplet
$\mathcal{T}=(\tau,\Psi^{(\tau)}_{ i},B_{ij}^{(\tau)},F_{ab}^{(\tau)-},\Lambda_{ i}^{(\tau)}, C^{(\tau)})  $
with weight $w=0$ and its anti-chiral partner $\overline{\mathcal{T}}=(\overline{\tau},\,{\Psi^{(\tau)i}},\,B^{(\tau)ij},\,F_{ab}^{(\tau)+},\,{\Lambda^{(\tau)i}},\,\overline{C}{}^{(\tau)})$.
We consider a one-dimensional profile of the coupling $\tau(x)$ as a function of $x$ and demand invariance under the $SO(4)$ subgroup of the $SO(5)$ isometry group.
In particular we have~$F^{(\tau)+}_{ab}=F^{(\tau)-}_{ab}=0$.

We wish to preserve the supersymmetry corresponding to the parameters given by~(\ref{epsilon-chi})-(\ref{eq:chi-zero-chirality}).
We set $\eta^i=\frac14 \gamma^\mu\nabla_\mu\epsilon^i$, $\eta_i=\frac14 \gamma^\mu\nabla_\mu\epsilon_i$.
For the coupling chiral multiplet, the conditions for supersymmetry
\begin{align}
\delta\Psi_{i}^{(\tau)} &= 
(\sla{\nabla}\tau)\,\e_{i}+\frac{1}{2}\,B^{(\tau)}_{ij}\,\e^{j}
=0\ , \quad
\label{eq:psi-equal-0}
\\
\delta\Lambda_{i}^{(\tau)} &=
-\frac{1}{2}\,\sla{\nabla}B^{(\tau)}_{ij}\ve^{jk}\e_{k}+\frac{1}{2}\,C^{(\tau)}\ve_{ij}\e^{j}
-B^{(\tau)}_{ij}\ve^{jk}\eta_{k}=0\ ,
\label{eq:lambda-equal-0}
\end{align}
determine $B_{ij}^{(\tau)}$ and $C^{(\tau)}$ to be given by
\begin{align}
B_{ij}^{(\tau)}=\frac{4\mathrm{i} { e^{{\rm i} \beta}}
\,r}{xf(x)}\tau'(x)\vec{n}\cdot\vec{\tau}_{ij}\ ,
\quad
C^{(\tau)}
=\frac{8{ e^{2{\rm i} \beta}}r^{2}}{x^{2}f(x)^{2}}\left(\tau''(x) - \frac{1}{x}\tau'(x)\right)\ .
\label{eq:coupling-multiplet-c}
\end{align}
Similarly, for the anti-chiral coupling multiplet, the conditions
\begin{align}
\delta\Psi^{i}&=
(\sla{\nabla}\bar{\tau})\,\e^{i}+\frac{1}{2}\,B^{(\tau)ij}\,\e_{j}
=0\ ,
\label{eq:delta-psi-equal-0-anti}
\\
\delta\Lambda^{i}&=
-\frac{1}{2}\,\sla{\nabla}B^{(\tau)\,ij}\ve_{jk}\e^{k}+\frac{1}{2}\,\bar{C}{}^{(\tau)}\ve^{ij}\e_{j}
-B^{(\tau)ij}\ve_{jk}\eta^{k}=0 \ , 
\label{eq:delta-antilambda-equal-0-anti}
\end{align}
whose expressions are related formally to~(\ref{eq:psi-equal-0}) and~(\ref{eq:lambda-equal-0}) by charge conjugation in Minkowski signature, lead to
\begin{align}
B^{(\tau)ij}=-\frac{\mathrm{i}\,{ e^{-{\rm i} \beta}}\,x}{rf(x)}\,\bar{\tau}'(x) \vec{n}\cdot\vec{\tau}^{\, ij}\ ,
\qquad
 \bar{C}^{(\tau)}
=\frac{{ e^{-2{\rm i} \beta}}\, x^{2}}{2r^{2}f(x)^{2}}\left(\bar{\tau}''(x)+\frac{3}{x}\,\bar{\tau}'(x)\right)\ .
\label{eq:coupling-multiplet-bc-bar}
\end{align}

To compare with the analysis in Section~\ref{sec:off-shell-flat}, let us introduce the variable $\theta$ via 
 \begin{equation}\label{x-theta}
 x=2r\tan\frac{\theta}{2}  \ .
\end{equation}
Then\begin{align}
B_{ij}^{(\tau)}
&=
\frac{2\, {\rm i}\,{ e^{{\rm i} \beta}} }{r} \cot(\theta/2)\,\frac{\d\tau}{\d\theta}\, \vec{n}\cdot\vec{\tau}_{ij} \ ,
\qquad
B^{(\tau)ij}=
-\frac{2\,{\rm i} \, { e^{-{\rm i} \beta}}  }{r}\,\tan(\theta/2)\,
\frac{\d\bar{\tau}}{\d\theta}\,
\vec{n}\cdot\vec{\tau}{}^{\, ij} \ , 
\label{B-round-theta}
\\
C^{(\tau)}
&=
\frac{{ e^{2{\rm i} \beta}}}{r^{2}}
\frac{\cos(\theta/2)}{\sin^3(\theta/2)} 
\left[
(\cos\theta-2)\,\frac{\d\tau}{\d\theta}
+\sin\theta\,\frac{\d^{2}\tau}{\d\theta^{2}}
\right]  \ , \\
\bar{C}{}^{(\tau)}
&=
\frac{{ e^{-2{\rm i} \beta}}}{r^{2}}
\frac{\sin(\theta/2)}{\cos^3(\theta/2)}
\left[
(\cos\theta+2)\,\frac{\d\bar{\tau}}{\d\theta}
+\sin\theta\,\frac{\d^{2}\bar{\tau}}{\d\theta^{2}}
\right]  \ .
\label{Cbar-round-theta}
\end{align}

We now take a limit to the step function profile
\begin{align}\label{tau-step-sphere}
\tau(\theta)
=
\begin{cases}
\tau_{+} & \text{for} \quad  0\leq \theta < \frac \pi 2\ ,
\\
\tau_{-} & \text{for} \quad  \frac \pi 2 <\theta  \leq \pi\ .
\end{cases}
\end{align}
We again set $\Delta\tau=\tau_+ - \tau_-$.
By applying the identities $x^n\delta(x)=0$ ($n\geq 1$), $x\,\delta'(x)=-\delta(x)$, $x^n\, \delta'(x)=0$ ($n\geq 2$), we get
\begin{equation} \label{BC-sphere-delta}
\begin{aligned}
B_{ij}^{(\tau)}
&=
- \frac{2\, {\rm i}\,{ e^{{\rm i} \beta}} }{r} \,
 \vec{n}\cdot\vec{\tau}_{ij} 
\,  \Delta\tau 
\,  \delta
  \left (\theta -  \frac \pi 2\right)
\ , &\qquad
B^{(\tau)ij} &=
\frac{2\,{\rm i}\,  { e^{-{\rm i} \beta}}  }{r}\,
\vec{n}\cdot\vec{\tau}{}^{\, ij}  \, \Delta\overline{\tau}\,
\delta\left(\theta-\frac \pi 2\right) 
\ ,
\\
C^{(\tau)}
&= - \frac{2\, { e^{2{\rm i}\beta} } }{r^2}\, \Delta\tau \, \delta'\left(\theta-\frac\pi2\right)
\ , &\qquad
\bar{C}{}^{(\tau)}
&=- \frac{2\, { e^{-2{\rm i}\beta} } }{r^2}\, \Delta\overline\tau \, \delta'\left(\theta-\frac\pi2\right)
\ .
\end{aligned}
\end{equation}
As we explain in Appendix~\ref{app:conf}, these expressions are related to the flat space results~(\ref{BC-flat-delta}) by the Weyl transformation, with the identification
 $\rho_{ij} = {\rm  i} \,{ e^{{\rm i}\beta}}\, \vec n\cdot\vec\tau_{ij}$, or equivalently $e^{{\rm i}\beta}=- {\rm i}\, e^{{\rm i}\alpha}$.
 
In Euclidean signature chiral and anti-chiral multiplets are independent.
Indeed for a generic profile $\tau(x)$, $(B^{(\tau)}_{ij},C^{(\tau)})$ and $(B^{(\tau)ij},\bar{C}{}^{(\tau)})$ as given in~(\ref{eq:coupling-multiplet-c}) and (\ref{eq:coupling-multiplet-bc-bar}) are not the complex conjugate of each other even though we demand that $\bar\tau(x)=\tau(x)^*$.
In the limit that the profile $\tau(x)$ becomes a step function, however,  $(B^{(\tau)}_{ij},C^{(\tau)})$ and $(B^{(\tau)ij},\bar{C}{}^{(\tau)})$ given in~(\ref{BC-sphere-delta}) are  the complex conjugate of each other.

Our construction involving a general profile $\tau(x)$ manifestly preserves~$OSp(2|2)_m$ at every step.
In the limit where~$\tau(x)$ becomes a step function~(\ref{tau-step-sphere}), the symmetry enhances, classically, to the full 3$d$ superconformal algebra $OSp(2|4)_{sc}$.
We regard a smooth profile as a UV regulator for the superconformal Janus interface on~$\BS^4$.

Repeating the analysis for the alternative choice~(\ref{eq:chi-zero-chirality-alternative}) leads to
\begin{equation}
\hspace{-4mm}
\text{(alternative)}\hspace{4mm}
\left\{
\begin{array}{ccl}
\vspace{1.5mm}
B_{ij}^{(\tau)}&=&
\displaystyle
-\frac{2\,{\rm i} \,  { e^{{\rm i} \beta}}   }{r}\,\tan(\theta/2)\,
\frac{\d\tau}{\d\theta}\,
\vec{n}\cdot\vec{\tau}_{ij}  \ ,
\
B^{(\tau)ij} 
=
\frac{2\,{\rm i} \, { e^{-{\rm i} \beta}}  }{r}\,\cot(\theta/2)\,
\frac{\d\bar{\tau}}{\d\theta}\,
\vec{n}\cdot\vec{\tau}{}^{\, ij} 
\ ,
\\
\vspace{1.5mm}
C^{(\tau)}
&=&
\displaystyle
\frac{{ e^{2{\rm i} \beta}} }{r^{2}}
\frac{\sin(\theta/2)}{\cos^3(\theta/2)} 
\left[
(\cos\theta+2)\,\frac{\d\tau}{\d\theta}
+\sin\theta\,\frac{\d^{2}\tau}{\d\theta^{2}}
\right] \ ,
\\
\bar{C}{}^{(\tau)}
&=&
\displaystyle
\frac{{ e^{-2{\rm i} \beta}} }{r^{2}}
\frac{\cos(\theta/2)}{\sin^3(\theta/2)}
\left[
(\cos\theta-2)\,\frac{\d\bar{\tau}}{\d\theta}
+\sin\theta\,\frac{\d^{2}\bar{\tau}}{\d\theta^{2}}
\right] \ .
\end{array} 
\right. 
\hspace{-6mm}
\end{equation}
These expressions are related to~(\ref{B-round-theta})-(\ref{Cbar-round-theta}) via $\theta\rightarrow\pi-\theta$.
In the limit~(\ref{tau-step-sphere}) they 
are related to~(\ref{BC-flat-delta}) with $\rho_{ij} =- {\rm  i} \,{ e^{{\rm i}\beta}}\, \vec n\cdot\vec\tau_{ij}$, or equivalently $e^{{\rm i}\beta}= + {\rm i}\, e^{{\rm i}\alpha}$,  by the Weyl transformation.

\subsection{Janus interface in gauge theory on $\mathbb{S}^{4}$}

In this section we review the general $\mathcal{N}=2$ superconformal gauge theory on $\BS^4$ and explain how to incorporate the half-BPS Janus interface that we constructed in Section~\ref{subsec:off-shell-construction-sphere} using the off-shell method.

A general $\mathcal{N}=2$ gauge theory involves a vector multiplet for a gauge group $G$ and matter hypermultiplets.
We allow $G$ to be a product of simple Lie groups and ignore the global structure because it plays no role for us.
Since we are interested in the conformal case, we assume that the hypermultiplets are in an appropriate representation of $G$ such that the beta functions for the gauge couplings exactly vanish.
As we will explain below, the hypermultiplets will enter our discussion only indirectly, and will be dropped for the most part.
To ease the notation we focus on a single gauge group factor with a complexified gauge coupling
\begin{equation}
\tau=\frac{\vartheta}{2\pi}+\frac{4\pi\,{\rm i}}{g_{\rm YM}^2} \ .
\end{equation}
Let~$\mathcal{V}
= ( X,\Omega_{i},A_{\mu},Y_{ij})$ be the corresponding vector multiplet.
In flat Euclidean space, the action is given as
\begin{align}\label{eq:action-vector-mult-flat}
\begin{aligned}
I_{\text{vector}}^\text{flat}
&=
\int \d^{4}x \,
{\rm Tr}
\Bigg[
\frac{1}{g_{\rm YM}^{2}}\Big(
4 D_{\mu}X  \, D^{\mu}\bar{X} 
-\frac{1}{2}\,\varepsilon^{ik}\,\varepsilon^{jl}\,Y_{ij}\,Y_{kl}
+2\bar{\Omega}_{i}\sla{D}\Omega^{i}
\\
&\qquad\qquad\qquad\qquad\qquad\qquad\qquad
+\frac{1}{2}\,F_{\mu\nu}F^{\mu\nu}
\Big)
+ \mathrm{i}\,
\frac{\vartheta}{16\pi^{2}}F_{\mu\nu}\tilde{F}^{\mu\nu}
\Bigg]\ .
\end{aligned}
\end{align}
Here ${\rm Tr}(\bullet\ \bullet)$ denotes an appropriately normalized inner product on the Lie algebra and reduces to the trace if $G=SU(N)$, and $D_\mu = \partial_\mu - {\rm i} A_\mu$ denotes the gauge covariant derivative.
We use hermitian generators $T_I$ and expand fields as~$X=T_I X^I$, $A_\mu = T_I A_\mu^I$, etc.  See Appendix~\ref{subsec:vector-multiplet}.
The dual field strength is defined as $\tilde{F}_{\mu\nu}= \frac12 \varepsilon_{\mu\nu}{}^{\rho\sigma} F_{\rho\sigma}$, where $\varepsilon_{\mu\nu}{}^{\rho\sigma}$ is the Levi-Civita tensor.
The action on the round sphere of radius $r$ can be obtained by a conformal transformation and is given as
\begin{align}\label{eq:action-vector-mult}
\begin{aligned}
I_{\text{vector}}
&=
\int \d^{4}x \, \sqrt{g} \,
{\rm Tr}
\Bigg[
\frac{1}{g_{\rm YM}^{2}}
\Big(
4 D_{\mu}X \, D^{\mu}\bar{X} + \frac{8}{r^2}  X\bar{X}
-\frac{1}{2}\,\varepsilon^{ik}\,\varepsilon^{jl}\,Y_{ij}\,Y_{kl}
+2\bar{\Omega}_{i}\sla{D}\Omega^{i}
\\
&\qquad\qquad\qquad\qquad\qquad\qquad\qquad\qquad\quad
+\frac{1}{2}\,F_{\mu\nu}F^{\mu\nu}
\Big)+ \mathrm{i}\,
\frac{\vartheta}{16\pi^{2}} F_{\mu\nu}\tilde{F}^{\mu\nu}
\Bigg]\ .
\end{aligned}
\end{align}
Here $g$ is the determinant of the metric.
To make this physical action positive semi-definite, as in~\cite{Pestun:2007rz,Hama:2012bg}, we impose the reality condition
\begin{equation} \label{Y-reality-Euc}
(Y^I_{ij})^* = - Y^{I ij} \quad  \left( \Longleftrightarrow (\vec Y^I)^*= -\vec Y^I \right)\ . 
\end{equation}
This is different from the physical reality condition in Minkowski signature.

The vector multiplet~$\mathcal{V}$ can be embedded into a chiral multiplet of Weyl weight~$w=1$, which we note as $\mathcal{A}(\mathcal{V})$, as
\begin{equation}
\begin{aligned}
A|_{\mathcal{A}({\mathcal{V}})} &=X\ , &
\quad
\Psi_{i}|_{\mathcal{A}({\mathcal{V}})} &=\Omega_{i}\ ,
&\quad
B_{ij}|_{\mathcal{A}({\mathcal{V}})}&=Y_{ij}\ ,
\\
F_{ab}^{-}|_{\mathcal{A}({\mathcal{V}})} &=
\frac12 \left(F_{ab}-\tilde{F}_{ab}\right)
\ , &\quad 
\Lambda_{i}|_{\mathcal{A}({\mathcal{V}})} &=-\varepsilon_{ij}\,\sla{D}\Omega^{j}\ ,
&\quad
C|_{\mathcal{A}({\mathcal{V}})} &=\left(-2D_\mu D^\mu + \frac{4}{r^2}\right)\overline{X}  \ .
\end{aligned}
\end{equation}
See Appendix~\ref{sec:N=2mult} for notations.

To introduce the Janus interface in gauge theory, we apply the construction of Section~\ref{subsec:off-shell-construction-sphere}.
We promote the gauge coupling constant $\tau$ to a position-dependent field $\tau(x)$, and further promote it
to the coupling chiral multiplet~$\mathcal{T}$ whose bottom component is~$\tau(x)$.
The coupling multiplet directly couples to the vector multiplet only; it affects the dynamics of hypermultiplets only indirectly through interactions involving the vector multiplet.
We also consider the anti-chiral multiplet whose bottom component is $\bar{\tau}(x)$ and denote it by $\bar{\mathcal{T}}$.
By using these multiplets, we can construct a SUSY invariant action as
\begin{align}\label{eq:action}
I_\text{Janus}
=
\frac{1}{8\pi\, \mathrm{i}}\int \d^{4}x\sqrt{g}\,
{\rm Tr} \left[C|_{\mathcal{T}\mathcal{A}({\mathcal{V}})^{2}}
-
\bar{C}|_{\bar{\mathcal{T}}\,\bar{\mathcal{A}}({\mathcal{V}}){}^{2}}
\right]\ ,
\end{align}
where $\mathcal{T}\mathcal{A}({\mathcal{V}})^{2}$ is the chiral multiplet constructed by the tensor calculus.
We give a short explanation for tensor calculus in Appendix~\ref{sec:tensor-calc} (with explicit formulas only given for the bosonic components).
For a constant profile $\tau(x)=\tau$, 
(\ref{eq:action}) reduces to the ordinary action for a vector multiplet \eqref{eq:action-vector-mult}:
\begin{equation}\label{sphere-action-C}
\frac{1}{8\pi\, \mathrm{i}}
\int \d^{4}x\sqrt{g}\,
{\rm Tr}  \left[
\tau\, C|_{\mathcal{A}({\mathcal{V}})^{2}}
-
\bar{\tau}\, \bar{C}|_{\bar{\mathcal{A}}({\mathcal{V})}^{2}}
\right]=I_{\text{vector}} \ .
\end{equation}

\section{SUSY localization, interface entropy, and Calabi's diastasis}\label{sec:diastasis}

In this section, we compute the sphere partition function in the presence of the Janus interface via SUSY localization.
We will study in detail only those aspects of localization which are affected by the Janus interface.

In the absence of an interface, the localization calculation proceeds in several steps that we sketch here~\cite{Pestun:2007rz}.
On top of the chirality condition~(\ref{eq:chi-zero-chirality}), one further constrains the SUSY parameters so that they generate an $SU(1|1)$ subalgebra~\cite{Gomis:2014woa}.
By supersymmetry, the path integral is invariant under the deformation of the physical action $I_\text{phys} \rightarrow I_\text{phys} + {\rm t} \delta V$, where~${\rm t}$ is a real deformation parameter,~$\delta$ is the supersymmetry variation, and $V$ is an appropriate fermionic functional of fields.
By taking the limit ${\rm t}\rightarrow \infty$, the path integral reduces to a sum over the saddle points of $\delta V$, or more precisely a finite-dimensional integral and a discrete infinite sum over the saddle point field configurations.
The saddle points are parametrized by~$a\in{\rm Lie}\, G$ and two non-negative integers $k$ and $\bar{k}$.
The variable $a$ parametrizes the so-called saddle point locus, which is the space of smooth saddle point configurations.
The integer $k$ parametrizes topologically non-trivial, zero-size instanton configurations localized at the north pole ($x=0$).
The integer $\bar{k}$ on the other hand parametrizes zero-size anti-instantons localized at the south pole ($x=\infty$).
In the absence of an interface, the partition function takes the form~\cite{Pestun:2007rz}
\begin{align}\label{eq:sphere-partition-function-integral}
Z_\text{SUSY}[\mathbb{S}^{4}](\tau, \bar{\tau})
=
\int [{\rm d}a]\,  e^{-I_{\text{cl}}(\tau,\bar{\tau})}Z_{\text{1-loop}}(a)
Z_{\text{inst}}(a,q)
Z_{\text{inst}}(a,\bar{q})
\ .
\end{align}
Here $I_{\text{cl}}$ is the classical action~(\ref{eq:action-vector-mult}) evaluated at the localization locus. 
$Z_\text{1-loop}(a)$ is the one-loop determinant that arise from the Gaussinan integration around the localization locus.
$Z_{\text{inst}}(a,q)=\sum_k q^k Z_k$ and $Z_{\text{inst}}(a,\bar{q})= \sum_{\bar{k}} \bar{q}^{\bar{k}}Z_{\bar k}$  are the instanton partition functions  with equivariant parameters $\epsilon_1=\epsilon_2=1/r$ and instanton counting parameters~$q=e^{2\pi \i\tau}$ and $\bar{q}=e^{-2\pi \i \overline{\tau}}$.
For details, we refer the reader to~\cite{Pestun:2007rz,Gomis:2011pf,Hama:2012bg,Pestun:2014mja}.

By the presence of an interface, the localization locus and the one-loop determinant are not affected because these are determined by $\delta V$ only.
But the value of the on-shell action~$I_\text{cl}$ and the instanton partition functions will be modified.

\subsection{On-shell action}\label{subsec:on-shell-action}
On the localization locus, the scalar field $X$ in the vector multiplet is constant.
We denote by $\mathcal{V}_{\text{cl}}$ the vector multiplet $\mathcal{V}$ evaluated at the localization locus.
It is given as%
\footnote{%
These are valid without imposing a chirality condition~(\ref{eq:chi-zero-chirality}) or~(\ref{eq:chi-zero-chirality-alternative}).}
\begin{align}\label{chiral-from-vec}
A|_{\mathcal{A}(\mathcal{V}_{\text{cl}})}=X\ ,
\qquad
B_{ij}|_{\mathcal{A}(\mathcal{V}_{\text{cl}})}=-\frac{2\,\mathrm{i} \,{ e^{{\rm i}\beta}}  X}{r}\, \vec{n}\cdot\vec{\tau}_{ij}\ ,
\qquad
C|_{\mathcal{A}(\mathcal{V}_{\text{cl}})}=\frac{4\, { e^{2{\rm i}\beta}}\,X}{r^{2}}\ .
\end{align}
From the tensor calculus rules given in Appendix \ref{sec:tensor-calc}, we can compute the components of the chiral multiplet $\mathcal{A}(\mathcal{V}_{\text{cl}})^{2}$ :
\begin{align}
A|_{\mathcal{A}(\mathcal{V}_{\text{cl}})^{2}}=X^{2}\ ,
\qquad
B_{ij}|_{\mathcal{A}(\mathcal{V}_{\text{cl}})^{2}}=-\frac{4\,\mathrm{i}\,  { e^{{\rm i}\beta}}\, X^{2}}{r}\,\vec{n}\cdot\vec{\tau}_{ij}\ ,
\qquad
C|_{\mathcal{A}(\mathcal{V}_{\text{cl}})^{2}}=\frac{12\,  { e^{2{\rm i}\beta}} \,X^{2}}{r^{2}}\ .
\label{eq:vector-multiplet-squared-classical}
\end{align}
Then we get 
\begin{equation}
\begin{aligned}
C|_{\mathcal{T}\mathcal{A}(\mathcal{V}_{\text{cl}})^{2}}
&=
\frac{12 \, { e^{2{\rm i}\beta} }\,
 X^2\,\tau(x)}{r^2} +X^2\,  C^{(\tau)}  + \frac{2\,{\rm i}\, { e^{{\rm i}\beta}} \,X^2}{r}\, \vec n \cdot \vec\tau^{\, ij}\, B_{ij}^{(\tau)} 
\\
&=
{ e^{2{\rm i}\beta} }
X^{2}\left[
\frac{12}{r^{2}}\,\tau(x)+
q^{(1)}(x)\,\tau'(x)+q^{(2)}(x)\,\tau''(x)
\right]\ ,
\label{eq:clasical-action-chiral}
\end{aligned}
\end{equation}
where
\begin{align}
q^{(1)}(x)=-\frac{8\,r^{2}}{x^{3}f(x)^{2}}
-\frac{16}{xf(x)}\ ,
\qquad
q^{(2)}
=
\frac{8\,r^{2}}{x^{2}f(x)^{2}}\ .
\end{align}
The chiral part of the classical action~\eqref{eq:action} is computed as
\begin{align}
\int \d^{4}x\, \sqrt{g}\, C|_{\mathcal{T}\mathcal{A}(\mathcal{V}_{\text{cl}})^{2}}
=
{2\pi^{2}}\int_{0}^{\infty} \d x\, x^{3}\,f^{4}\, C|_{\mathcal{T}\mathcal{A}(\mathcal{V}_{\text{cl}})^{2}}
=
{32\pi^{2}}\, { e^{2{\rm i}\beta}}\, X^{2}\,r^{2}\,\tau(0)\ .
\end{align}
A similar computation can be done for the anti-chiral part using
\begin{equation}\label{anti-chiral-from-vec}
\overline{A}|_{\overline{\mathcal{A}}(\mathcal{V}_\text{cl})}= \overline{X} \ , \quad
B^{ij}|_{\overline{\mathcal{A}}(\mathcal{V}_{\text{cl}})} =
-\frac{2\,\mathrm{i}\, { e^{-{\rm i}\beta}}\, \overline{X} }{r}\, \vec{n}\cdot\vec{\tau}^{\, ij}\ ,
\qquad
\overline{C}|_{\overline{\mathcal{A}}(\mathcal{V}_{\text{cl}})}=
\frac{4\, { e^{-2{\rm i}\beta}}\,\overline{X}}{r^{2}}\ .
\end{equation}
We obtain
\begin{align}
\int \d^{4}x\, \sqrt{g}\, C|_{\overline{\mathcal{T}}\overline{\mathcal{A}}(\mathcal{V}_{\text{cl}})^{2}}
=
{32\pi^{2}} { e^{-2{\rm i}\beta}} \,\overline{X}^{2}r^{2}\,\tau(\infty)\ .
\end{align}

For the chiral and anti-chiral multiplets that arise from a single vector multiplet, $B_{ij}$ and $B^{ij}$ are related: $B_{ij}|_{\mathcal{A}(\mathcal{V}_{\text{cl}})}= \vec Y \cdot \vec\tau_{ij}$, $B^{ij}|_{\overline{\mathcal{A}}(\mathcal{V}_{\text{cl}})}= \vec Y \cdot \vec\tau^{\,ij}$.  In Euclidean signature the vector~$\vec Y$ is pure imaginary rather than real.  See~(\ref{Y-reality-Euc}).
Comparing~(\ref{chiral-from-vec}) and~(\ref{anti-chiral-from-vec}) we can write
\begin{equation}
X= \frac12{ e^{-{\rm i}\beta}}\, a
 \ , \qquad
\overline{X}=   \frac12 { e^{{\rm i}\beta}}\, a
\end{equation}
with $a$ real.
The normalization for~$a$ is chosen to be consistent with~\cite{Pestun:2007rz}.

The on-shell value of the action~(\ref{eq:action}) is the sum of the chiral and anti-chiral parts
\begin{align}\label{eq:classical-action-janus}
I_\text{Janus}
= - \mathrm{i}\,\pi\, r^{2}\,(\tau_+ - \bar{\tau}{}_- )\,{\rm Tr}\, a^{2}\ ,
\end{align}
where $\tau_+\equiv\tau(0)$ and $\tau_-\equiv\tau(\infty)$.
This result is related to the classical action without the interface by analytically continuing $(\tau,\bar{\tau})$  to $(\tau_+,\bar{\tau}{}_-)$:
\begin{align}
I_\text{Janus}
=
I_{\text{cl}}(\tau_+,\bar{\tau}{}_-)\ .
\end{align}

\subsection{Instanton partition functions}

The instanton partition functions without the Janus interface in \eqref{eq:sphere-partition-function-integral} arise from the fluctuation modes around the instantons and the anti-instantons localized at the north and south poles, respectively.
These localized topological excitations contribute to the physical action~(\ref{sphere-action-C}) and yield the weights $q^k$ and $\overline{q}^{\bar{k}}$.
In the presence of the Janus interface, the weights are modified to $q_+^k$ and $\overline{q}_-^{\bar{k}}$, where $q_+=e^{2\pi {\rm i} \tau_+}$, $\bar q_-=e^{-2\pi {\rm i} \bar\tau_-}$.
In other words, the Janus interface induces an analytic continuation of the instanton partition functions~$(\tau,\bar{\tau}) \rightarrow (\tau_+,\bar{\tau}{}_-)$.

Thus in the expression~(\ref{eq:sphere-partition-function-integral}), $I_\text{cl}(\tau,\bar\tau)$, $Z_{\text{inst}}(a,q)$, and $Z_{\text{inst}}(a,\bar{q})$ are replaced by~$I_\text{cl}(\tau_+,\bar\tau_-)$, $Z_{\text{inst}}(a,q_+)$, and $Z_{\text{inst}}(a,\bar{q}_-)$, respectively.
We assume that at least when the difference between $\tau_+$ and $\tau_-$ is small enough, the integral in~(\ref{eq:sphere-partition-function-integral}) remains convergent with the contours of integration suitably chosen.
Then the whole partition function in the presence of the Janus interface is given by the analytic continuation~$(\tau,\bar{\tau}) \rightarrow (\tau_+,\bar{\tau}{}_-)$.

\subsection{K\"ahler ambiguity and finite counterterms}\label{sec:counterterm}

SUSY localization computes the partition function in a specific renormalization scheme.
Other schemes are possible, and two different schemes are related by a finite counterterm.
As shown in~\cite{Gerchkovitz:2014gta} for 4$d$ $\mathcal{N}=2$ superconformal field theories coupled to an off-shell Poincar\'e supergravity, a renormalization scheme corresponds to a particular choice of the K\"ahler potential on the conformal manifold.
Two choices are related by a K\"ahler transformation, which corresponds to a finite supergravity counterterm~\cite{,Gomis:2014woa}. In this section, we evaluate this counterterm in the presence of the Janus interface.

The relevant off-shell Poincar\'e supergravity is obtained by gauge fixing conformal supergravity using compensating multiplets.
One of the compensators is the vector multitplet~$\mathcal{V}_{\text{c}}$ whose components take values~\cite{,Gomis:2014woa}%
\footnote{%
We note that~$Y_{ij}|_{{\mathcal{V}}_{\text{c}}}$ and $Y^{ij}|_{{\mathcal{V}}_{\text{c}}}$ violate the physical reality condition: $(Y_{ij}|_{{\mathcal{V}}_{\text{c}}})^* \neq Y^{ij}|_{{\mathcal{V}}_{\text{c}}}$
}
\begin{align}
 {X}|_{{\mathcal{V}}_{\text{c}}} &={\mu}e^{-\mathrm{i}\beta}\  ,
&\quad
Y_{ij}|_{{\mathcal{V}}_{\text{c}}} &=-\frac{2\,\i\,{\mu}}{r}\,(\vec{n}\cdot\vec{\tau})_{ij} \ ,
&\quad
{\Omega}_{i}|_{{\mathcal{V}}_{\text{c}}} &=F^{-}_{\mu\nu}|_{{\mathcal{V}}_{\text{c}}}=0 \ ,\label{Vc1} \\
\bar{X}|_{{\mathcal{V}}_{\text{c}}} &=\mu e^{+\mathrm{i}\beta}\ ,
&\quad
Y^{ij}|_{{\mathcal{V}}_{\text{c}}}
&=-\frac{2\,\i\,{\mu}}{r}\,(\vec{n}\cdot\vec{\tau})^{ij}\  ,
&\quad
\bar{\Omega}{}^{i}|_{{\mathcal{V}}_{\text{c}}}
&=F^{+}_{\mu\nu}|_{{\mathcal{V}}_{\text{c}}}
=0 \ , \label{Vc2}
\end{align}
where $\mu>0$ is an arbitrary mass scale.
This vector multiplet can be embedded into the
anti-chiral multiplet
 $\overline{\Phi}:=\overline{\mathcal{A}}(\mathcal{V}_{\text{c}})$ with Weyl weight one. 
We can further construct a chiral multiplet $\mathbb{T}(\log\bar{\Phi})$ with Weyl weight two from $\overline{\Phi}$.%
\footnote{%
In flat space, with $\overline{\Phi}$ viewed as an anti-chiral {\it superfield}, the top component of~$\log\overline{\Phi}$ is a chiral primary of Weyl weight 2~\cite{Butter:2013lta}.
A chiral multiplet can be constructed by repeated SUSY transformations such that the chiral primary is its bottom component.
$\mathbb{T}(\log\bar{\Phi})$ is the curved version of this chiral multiplet.
}
Its components are given by
\begin{align}
A|_{\mathbb{T}(\log\bar{\Phi})}
&=
\frac{2\,{e^{-2\mathrm{i}\beta}}}{r^{2}} ,
\\
B_{ij}|_{\mathbb{T}(\log\bar{\Phi})}
&=-\frac{8\,\mathrm{i}\,{e^{-\mathrm{i}\beta}}}{r^{3}}\,(\vec{n}\cdot\vec{\tau})_{ij} ,
\\
C|_{\mathbb{T}(\log\bar{\Phi})}
&=
{\frac{24}{r^{4}}} .
\end{align}

Next, we compute the components of $\mathcal{F}(\mathcal{T})$ for an arbitrary holomorphic function $\mathcal{F}(\cdot)$ via the tensor calculus rules given in Appendix~\ref{sec:tensor-calc}.
Its components are given by
\begin{align}
A|_{\mathcal{F}(\mathcal{T})} &=\mathcal{F}(\tau) \ ,
\\
B_{ij}|_{\mathcal{F}(\mathcal{T})}
&=
\frac{\d\mathcal{F}(\tau)}{\d x}\,\frac{\mathrm{i}\,r\,{e^{\mathrm{i}\beta}}}{xf(x)}\,(\vec{n}\cdot\vec{\tau})_{ij} \ ,
\\
C|_{\mathcal{F}(\mathcal{T})}
&=
\frac{8\,r^{2}\,{e^{2\mathrm{i}\beta}}}{x^{2}f^{2}}\left(
\frac{\d^{2}\mathcal{F}(\tau)}{\d x^{2}}
-\frac{1}{x}
\frac{\d\mathcal{F}(\tau)}{\d x}
\right) \ .
\end{align}

The SUSY invariant counterterm considered in~\cite{,Gomis:2014woa} is the top component of the product chiral multiplet $\mathcal{F}(\mathcal{T}) \mathbb{T}(\log\bar{\Phi})$.
It can be computed by the tensor calculus rules given in Appeneix~\ref{sec:tensor-calc}.
Note that the components of $\mathcal{F}(\mathcal{T})$ are obtained from those of the coupling multiplet $\mathcal{T}$ given in~\eqref{eq:coupling-multiplet-c}  by replacing $\tau$ with $\mathcal{F}(\tau)$.
Similarly the components of $\mathbb{T}(\log\bar{\Phi})$ are obtained from those of $\mathcal{A}(\mathcal{V}_{\text{cl}})^{2}$ given in~\eqref{eq:vector-multiplet-squared-classical} by replacing $X^{2}$ with $\frac{2{e^{-2\mathrm{i}\beta}}}{r^{2}}$.
Therefore the top component of $\mathcal{F}(\mathcal{T}) \mathbb{T}(\log\bar{\Phi})$ can be obtained from $C|_{\mathcal{T}\mathcal{A}(\mathcal{V}_{\text{cl}})^{2}}$ in~(\ref{eq:clasical-action-chiral}) by the same substitutions:
\begin{equation}
C\big|_{\mathcal{F}(\mathcal{T})\mathbb{T}(\log\bar{\Phi})}
=\frac{2}{r^{2}}
\left[
\frac{2}{r^{2}}
\mathcal{F}(\tau)
+q^{(1)}(x)\,\frac{\d\mathcal{F}(\tau)}{\d x}
+q^{(2)}(x)\,\frac{\d^{2}\mathcal{F}(\tau)}{\d x^{2}}
\right] \ .
\end{equation}
Thus 
\begin{align}
\int \d^{4}x\,\sqrt{g}\,C|_{\mathcal{F}(\mathcal{T})\mathbb{T}(\log\bar{\Phi})}
=
64\pi^{2}\,\mathcal{F}(\tau_+)\ .
\end{align}
Similarly we can compute the anti-chiral counterterm constructed from the anti-chiral coupling multiplet~$\bar{\mathcal{T}}$ and the compensating vector multiplet $\mathcal{V}_{\text{c}}$:
\begin{align}
\int \d^{4}x\,\sqrt{g}\,\bar{C}|_{\bar{\mathcal{F}}(\bar{\mathcal{T}})\mathbb{T}(\log{\Phi})}
=64\pi^{2}\,\bar{\mathcal{F}}(\bar{\tau}{}_-)\ .
\end{align}
The anti-holomorphic $\bar{\mathcal{F}}(\bar\tau)$ is the complex conjugate of the holomorphic function $\mathcal{F}(\tau)$ when~$\bar\tau=\tau^*$.

\subsection{Interface entropy as Calabi's diastasis}\label{subsec:partition-function-and-interface-entropy}

By assembling the results above, we now relate the sphere partition function in the presence of the Janus interface to Calabi's diastasis.
By a previous result~\cite{Gerchkovitz:2014gta} the sphere partition function in the absence of the Janus interface can be written as
\begin{align}\label{eq:sphere-partition-function}
Z_\text{SUSY}[\mathbb{S}^{4}](\tau, \bar{\tau})=e^{K(\tau, \bar{\tau})/12}\ .
\end{align}
We saw that the sphere partition function with the Janus interface can be obtained by analytically continuing $(\tau,\bar{\tau}) \to (\tau_+,\bar{\tau}{}_-)$ in the sphere partition function \eqref{eq:sphere-partition-function-integral}. Then by using \eqref{eq:sphere-partition-function} we can write the sphere partition function in the presence of the Janus interface in terms of the analytically continued K\"{a}hler potential as follows:
\begin{align}\label{eq:partition-function-janus}
Z^{\mathcal{I}}_\text{SUSY}[\mathbb{S}^{4}]
=
e^{K(\tau_+, \bar{\tau}{}_-)/12}\ .
\end{align}
Besides we can add the counterterms constructed in the previous section to the action.
These terms modify the sphere partition function.
With proper normalizations this modification is (an analytically continued version of) the K\"ahler transformation
\begin{align}\label{eq:kahler-transformation}
K(\tau_+,\bar{\tau}{}_-)
\to
K(\tau_+,\bar{\tau}{}_-)
+\mathcal{F}(\tau_+)
+\bar{\mathcal{F}}(\bar{\tau}{}_-)\ .
\end{align}

Then by substituting the result \eqref{eq:partition-function-janus} into \eqref{eq:relation-ie-partition-function}, we conclude that the interface entropy can be written in terms of the analytically continued K\"{a}hler potentials as
\begin{align}\label{eq:janus-interface-entropy}
S_\CI
=-\frac{1}{24}
\left[
K(\tau_+, \bar{\tau}{}_+)+K(\tau_-, \bar{\tau}{}_-)
-
K(\tau_+, \bar{\tau}{}_-)-K(\tau_-, \bar{\tau}{}_+)
\right]\ .
\end{align}
The combination in the bracket is Calabi's diastasis~\eqref{diastasis-def} defined in the introduction.
Calabi's diastasis~\eqref{diastasis-def} and the entropy of the Janus interface~\eqref{eq:janus-interface-entropy} is invariant under the transformation~\eqref{eq:kahler-transformation}.

\section{A holographic example}\label{sec:holography}


$\CN=4$ supersymmetric Yang-Mills theory with the maximally supersymmetric conformal interface has a dual gravity description by the supersymmetric Janus solution in the type IIB supergravity \cite{DHoker:2007zhm}.
The solution respects $SO(1,4) \times SO(3) \times SO(3)$ symmetry associated with the conformal symmetry on the three-dimensional interface and the unbroken $R$-symmetry.
The metric takes the form
\begin{align}\label{SUSY_Janus}
	\d s^2 = f_4^2\, \d s_{AdS_4}^2 + \rho^2\, \d v \d \bar v + f_1^2\, \d s_{\BS^2}^2 + f_2^2\, \d s_{\BS^2}^2 \ ,
\end{align}
where $\d s_{\BS^2}^2$ is the metric of a unit 2-sphere and $v = x + \i y$ is a complex coordinate on a strip with the ranges $ x \in \BR$ and $0\le y \le \pi/2$.
The functions $f_4, \rho, f_1, f_2$ are determined by two real functions $h_1(v, \bar v)$ and $h_2(v, \bar v)$ as 
\begin{align}
	\begin{aligned}
		f_4^8 &= 16\,\frac{F_1 F_2}{W^2} \ , &\qquad   \rho^8 &= \frac{2^8F_1 F_2 W^2}{h_1^4 h_2^4} \ , \\
		f_1^8 &= 16 h_1^8\,\frac{F_2 W^2}{F_1^3} \ , &\qquad   f_2^8 &=  16 h_2^8\,\frac{F_1 W^2}{F_2^3}  \ , 
	\end{aligned}
\end{align}
where 
\begin{align}
	F_i = 2h_1 h_2 |\partial_v h_i|^2 - h_i^2 W \quad (i=1,2) \ , \qquad W = \partial_v \partial_{\bar v} (h_1 h_2) \ .
\end{align}
The real functions are given by
\begin{align}
	h_1(v, \bar v) = -\i\, \alpha_1\sinh \left( v - \frac{\Delta\phi}{2}\right) + \text{c.c.} \ , \qquad h_2(v, \bar v) = \alpha_2\cosh \left( v + \frac{\Delta\phi}{2}\right) + \text{c.c.}\ .
\end{align}
This solution has two asymptotic regions at $x\to \pm \infty$ corresponding to the two sides of the Janus interface.
The real parameters $\alpha_1, \alpha_2$ and $\Delta\phi$ fix the AdS radius $L$ and the Yang-Mills couplings $g_\text{YM}^\pm$ by the relations:
\begin{align}
	L^4 = 16 |\alpha_1 \alpha_2|\cosh \Delta\phi \ , \qquad (g_\text{YM}^\pm)^2 = 4\pi \bigg| \frac{\alpha_2}{\alpha_1}\bigg|\, e^{\pm \Delta\phi} \ .
\end{align}

\subsection{Sphere free energy}
We are interested in the sphere free energy of the interface CFT dual to the SUSY Janus solution.
It can be calculated holographically by evaluating the on-shell action after a consistent truncation to four dimensions \cite{Assel:2012cp}:
\begin{align}
	I = - \frac{3\cdot 2^6\, \text{Vol}(\BS^2)^2}{16\pi G_N} \int_{AdS_4} \d^4 x\sqrt{g_{(4)}}\int \d x\, \d y\, W h_1 h_2 \ ,
\end{align}
where $G_N$ is the Newton constant in ten dimensions.
In terms of the coordinate~$\lambda$ such that 
\begin{align}
	\d s^2_{AdS_4} = \frac{1}{\cos^2\lambda}\left[ \d \lambda^2 + \sin^2\lambda\,\d s^2_{\BS^3}\right] \ ,
\end{align}
with $0\le \lambda \le \pi/2$, the integral becomes
\begin{align}
	I = \frac{3\, \text{Vol}(\BS^2)^2\,\text{Vol}(\BS^3)\,L^8}{2^6\pi G_N} \int_0^{\pi/2} \d\lambda \,\frac{\sin^3 \lambda}{\cos^4\lambda} \,\int_0^{\pi/2}\d y\sin^2 (2y)\,\int_{-\infty}^\infty \d x\left( 1 + \frac{\cosh (2x)}{\cosh (\Delta\phi)}\right) \ .
\end{align}
This is divergent and requires a cutoff.

To regularize the integral, we adopt the single cutoff procedure \cite{Bak:2016rpn,Gutperle:2016gfe},%
\footnote{%
There are other cutoff procedures for regularization in Janus geometry \cite{Estes:2014hka,Gutperle:2016gfe}.}
 which cuts out the spacetime outside the UV boundary hypersurface satisfying
\begin{align}
	\frac{f_4}{Z} = \frac{L}{\delta} \ ,\qquad Z\equiv \cos\lambda \ .
\end{align}
Then the integration for $x$ is restricted from $x_-(Z, y)$ to $x_+(Z, y)$ defined by $f_4(x_\pm) = L Z/\delta$ for $Z$ fixed.
It also restricts the range of $Z$ from $Z_\ast \equiv f_4(0)\delta/L$ to $1$.
We can perform the integration over $x$ by expanding $x_\pm$ in $\delta/Z$:%
\footnote{%
This expansion differs from (3.10) in \cite{Gutperle:2016gfe}.}
\begin{align}\label{Cutoff_x}
	x_\pm(Z, y) = \pm\frac{1}{2}\log \left( 4 \cosh(\Delta\phi)\,\frac{Z^2}{\delta^2}\right) - \frac{ \cos(2y)\,\tanh(\Delta\phi)\pm 2}{8}   \left( \frac{\delta}{Z}\right)^2 +\CO\left(\frac{\delta^4}{Z^4}\right) \ ,
\end{align}
It follows that the integral over $x$ becomes
\begin{align}
	\int_{x_-(Z,y)}^{x_+(Z,y)} \d x\,\left( 1 + \frac{\cosh (2x)}{\cosh (\Delta\phi)}\right) = \log \left( 4 \cosh(\Delta\phi)\,\frac{Z^2}{\delta^2}\right) + 2\frac{Z^2}{\delta^2} - 1 + \CO\left(\frac{\delta^2}{Z^2}\right) \ .
\end{align}
Hence the regularized on-shell action becomes
\begin{align}
	\begin{aligned}
		I &= \frac{3\, \text{Vol}(\BS^2)^2\,\text{Vol}(\BS^3)\,L^8}{2^6\pi G_N} \,\int_0^{\pi/2}\d y\,\sin^2(2y) \int_{Z_\ast}^1\d Z\,\frac{1-Z^2}{Z^4}
		\\
		&
		\qquad\qquad\qquad\qquad\qquad\qquad
		\times \left[ \log \left( 4 \cosh(\Delta\phi)\,\frac{Z^2}{\delta^2}\right) + 2\frac{Z^2}{\delta^2} + 1 + \CO\left(\frac{\delta^2}{Z^2}\right)\right] \\
			&= \frac{\text{Vol}(\BS^2)^2\,\text{Vol}(\BS^3)\,L^8}{2^{7} G_N} \,\left[ \frac{c_3}{\delta^3} + \frac{c_2}{\delta^2} +  \frac{c_1}{\delta} + \log\left( \frac{4 \cosh(\Delta\phi)}{\delta^2}\right) + \frac{5}{3} + \CO(\delta^2)\right]\ ,
	\end{aligned}
\end{align}
where we do not bother to write down the coefficients $c_i~(i=1,2,3)$ which contain logarithmically divergent terms.
Subtracting the bulk contribution, the universal part of the free energy is 
\begin{align}
	\Delta I = I - I|_{\Delta\phi = 0} = \frac{\text{Vol}(\BS^2)^2\,\text{Vol}(\BS^3)\,L^8}{2^{7} G_N}\log \cosh (\Delta\phi) \ .
\end{align}
Using the relation of the Newton constant and the rank $N$ of the gauge group
\begin{align}
	G_N = \frac{\text{Vol}(\BS^2)^2\,\text{Vol}(\BS^3)\,L^8}{2^6 N^2} \ ,
\end{align}
we find the sphere free energy of the supersymmetric Janus solution of the form
\begin{align}\label{Janus_SFE}
	\Delta I = \frac{N^2}{2}\log \left[ 1 + \frac{(g_\text{YM}^+ - g_\text{YM}^- )^2}{2g_\text{YM}^+ g_\text{YM}^-}\right] \ ,
\end{align}
which is minus the interface entropy obtained in \cite{Estes:2014hka}.
This is in accordance with the universal relation between the sphere free energy and entanglement entropy across a sphere in ICFT~\cite{Kobayashi:2018lil}.

Applying an $SL(2,\BR)$ transformation of the type IIB supergravity on the Janus solution without a theta-angle generates a new solution with a complexified coupling 
\begin{align}
	\tau = \frac{\vartheta}{2\pi} + \frac{4\pi \i}{g_\text{YM}^2} \ ,
\end{align}
jumping across an interface.
Hence the universal part of the sphere free energy of the supersymmetric Janus solution with the coupling taking values~$\tau_\pm$ across an interface is \cite{Goto:2018zrp}
\begin{align}
	\Delta I = \frac{1}{24}\left[ K(\tau_+, \bar{\tau}_+) + K(\tau_-, \bar{\tau}_-) - K(\tau_+, \bar{\tau}_-) - K(\tau_-, \bar{\tau}_+)\right] \ ,
\end{align}
where $K$ is the K\"ahler potential given by
\begin{align}
	K(\tau, \bar \tau) = -6N^2\log \left[ \i \,(\bar \tau - \tau)\right] \ .
\end{align}
If we identify the holographic free energy with the sphere partition function by the relation
\begin{align}
	\Delta I = - \log \frac{Z^{(\text{ICFT})}[\BS^4]}{(Z^{(\text{CFT}_+)}[\BS^4] \, Z^{(\text{CFT}_-)}[\BS^4] )^{1/2}} \ ,
\end{align}
we find the sphere partition function 
\begin{align}
	Z^{(\text{ICFT})}[\BS^4](\tau_+, \bar\tau_-) \propto \big| e^{K(\tau_+, \bar\tau_-)/12} \big| \ ,
\end{align}
which is consistent with our assumption~(\ref{eq:assumption}).

\subsection{Entanglement entropy}

Next we consider the entanglement entropy across a sphere centered at the origin of the Janus interface.
In the holographic system described by the metric \eqref{SUSY_Janus} it is convenient to use the Poincar\'e coordinates of the Lorentzian AdS spacetime, in terms of which the metric is
\begin{align}
	\d s^2_{AdS_4} = \frac{1}{z^2}\left[ \d z^2 - \d t^2 + \d r^2 + r^2\, \d \phi^2 \right] \ .	
\end{align}
The spherical entangling surface is on the boundary at a constant time slice
\begin{align}
	\Sigma = \{ t = 0, \, r = R , \, z=0\} \ .
\end{align}
The holographic entanglement entropy is given by the area of the minimal surface anchored on $\Sigma$
\cite{Ryu:2006bv,Ryu:2006ef},  
\begin{align}\label{EE_Janus}
	S = \frac{\text{Vol}(\BS^2)^2\,\text{Vol}(\BS^1)}{4G_N} \, \left(\int \d x\, \d y\,( f_1\, f_2\,f_4\, \rho)^2\right) \int \d z \,\frac{r}{z^2}\sqrt{1 + (\partial_z r)^2} \ ,
\end{align}
where the minimal surface is determined by a function $r(z)$ which is independent of $(t,\phi)$ due to the spherical symmetry.
Varying the area functional with respect to $r(z)$ yields the equation of motion, which turns out to allow for a simple solution \cite{Jensen:2013lxa}
\begin{align}
	r = \sqrt{R^2 - z^2} \ .
\end{align}
To evaluate the entropy \eqref{EE_Janus} on shell, we need a regularization for the UV divergence.
In the single cutoff prescription we cut out the spacetime by the UV boundary hypersurface%
\footnote{%
The UV regulator $\varepsilon$ is different from $\delta$ used for the free energy calculation.
It is not clear how to relate them as $\varepsilon$ and $\delta$ are introduced for the Lorentzian and the Euclidean spacetimes, respectively.
}
\begin{align}
	\frac{f_4}{z} = \frac{L}{\varepsilon} \ ,
\end{align}
which restricts the integration range for $x$ to $x_- (z, y) \le x \le x_+ (z, y)$ with $x_\pm(z,y)$ given by~\eqref{Cutoff_x}, where $(z, \delta)$ are replaced with $(z,\varepsilon)$.
Also the $z$ integral is restricted to $z_\ast \equiv f_4(0)\varepsilon/L \leq z \leq R$.
The regularized expression of the entropy becomes
\begin{align}
	S = \frac{2^4\pi\,\text{Vol}(\BS^2)^2\,L^8}{2^4G_N} \,\int_0^{\pi/2}\d y\, \sin^2 (2y)\,\int_{z_\ast/R}^1 \, \frac{\d z}{z^2}\, \int_{x_-(z, y)}^{x_+(z,y)} \d x\, \left( 1 + \frac{\cosh (2x)}{\cosh (\Delta\phi)}\right) \ .
\end{align}
Repeating the same type of the calculation as for the free energy, we find the universal part of the interface entropy
\begin{align}
	S_\CI|_\text{univ} = - \frac{N^2}{2}\log \cosh(\Delta\phi) \ ,
\end{align}
which agrees with the result obtained using another regularization \cite{Estes:2014hka}.
We note that the interface entropy is minus the sphere free energy as expected from the CFT consideration, {\it i.e.}, from the relation~(\ref{eq:relation-ie-partition-function-nonSUSY-intro}).

\section{Discussion}\label{sec:discussion}

\subsection{Super-Weyl anomaly}\label{sec:super-Weyl}

In 2$d$ with $\mathcal{N}=(2,2)$ SUSY one can use the super-Weyl anomaly of~\cite{Bachas:2016bzn} to prove the 2$d$ and boundary ($\mathcal{B}$) version of the relation~(\ref{eq:assumption}), {\it i.e.}, $Z^\text{(BCFT)}[\BS^2] = \left|Z^\mathcal{B}_\text{SUSY}[\BS^2] \right|$.
Indeed
$Z^\text{(BCFT)}[\BS^2]$ is the overlap of the boundary state and the ground state in the NSNS sector.
This overlap is nothing but the $g$-factor, which was shown to be a boundary contribution to the entanglement entropy in~\cite{Calabrese:2004eu}.
The NSNS overlap on the other hand was shown to be the absolute value of the SUSY partition function in the presence of a boundary in~\cite{Bachas:2016bzn} using the super-Weyl anomaly.

Somewhat more explicitly, on a half-plane~$x^1\leq 0$ and in Euclidean signature, the super-Weyl variation of the logarithm of the partition function reads, in superconformal gauge,
\begin{equation}
\delta_\Sigma \log Z \supset \delta\left[
-\frac{1}{4\pi}\int {\rm d}^2x \left(\Box(\sigma-\i\, a)h^\Omega+\Box(\sigma+\i\, a)\bar h{}^\Omega\right) +\frac{\rm i}{4\pi}\int {\rm d}x^2 (\bar w h^\Omega - w \bar h{}^\Omega)
\right] \ .
\end{equation}
See~\cite{Bachas:2016bzn} for notations.
The inside of the large bracket is essentially~$ \log Z$.
The twisted chiral superfield~$\Sigma=\sigma+ \i \,a +\theta^+\bar\chi_++\bar\theta{}^-\chi_- + \theta^+ \bar\theta{}^- w$ is the supersymmetric version of the Weyl factor $\sigma$ that represents the metric $g_{\mu\nu}=e^{2\sigma}\delta_{\mu\nu}$ in the conformal gauge.
For the round sphere $\sigma=-\log(1+|z|^2)$, where $z=x^1+{\rm i}x^2$.
If one demands supersymmetry used for localization but gives up conformal invariance, we get~$w=\bar w= -2\i/(1+|z|^2)$ and $a=0$.%
\footnote{%
The values of $w$ and $\bar w$ violate unitarity~\cite{Closset:2014pda}.}
This gives the supersymmetric hemisphere partition function~\cite{Sugishita:2013jca,Honda:2013uca,Hori:2013ika} as $Z_\text{SUSY}^\mathcal{I}[\BS^2] \sim \exp h^\Omega$.
If one demands conformal invariance we get~$w=\bar w = a=0$.
This gives~$Z^\text{(ICFT)}[\BS^2]\sim \exp \frac12 (h^\Omega+\bar h{}^\Omega)$.
We thus have~$Z^\text{(ICFT)}[\BS^2] = \left|Z^\mathcal{I}_\text{SUSY}[\BS^2] \right|$.
This explanation is similar in spirit to~\cite{Closset:2012vg}.

It would be nice to extend the analysis of~\cite{Bachas:2016bzn} to 4$d$.

\subsection{Complex partition functions and a Chern-Simons counterterm}

For 3$d$ $\mathcal{N}=2$ superconformal field theories, a relation similar to~(\ref{eq:assumption}), $ Z^\text{(CFT)}[\BS^3] = \left|Z_\text{SUSY}[\BS^3] \right|$, was shown using a supersymmetric Chern-Simons coupling as follows~\cite{Closset:2012vg}.
The conformal partition function~$ Z^\text{(CFT)}[\BS^3]$ is defined in a conformally invariant renormalization scheme and is real and positive.
The supersymmetric partition function~$Z_\text{SUSY}[\BS^3]$ is computed by SUSY localization in some renormalization scheme and is complex.
The two schemes and the two partition functions should differ by finite counterterms.
The relevant counterterm is the $Z$-$Z$ Chern-Simons term constructed from the off-shell Poincar\'e supergravity multiplet.
It violates conformal invariance, and involves a field~$H$ which in the supersymmetric $\BS^3$ background takes a value that violates unitarity.
The on-shell value of the $Z$-$Z$ Chern-Simons term is pure imaginary, and is responsible for making~$Z_\text{SUSY}[\BS^3]$ complex.

We expect that an essentially identical explanation should be possible.
Indeed in the extreme case that the bulk 4$d$ $\mathcal{N}=2$ superconformal theory on $\BS^4$ is trivial, a half-BPS interface is nothing but a 3$d$ superconformal field theory living on~$\BS^3$.

It seems plausible that the assumption~(\ref{eq:assumption}) can be shown along the following line.
One can impose boundary conditions on symmetry parameters in a way similar to~\cite{Belyaev:2008ex} so that the 4$d$ $\mathcal{N}=2$ Weyl multiplet restricted to a 3$d$ boundary decomposes into 3$d$ $\mathcal{N}=2$ multiplets.
The restricted 4$d$ Weyl multiplet would include the 3$d$ Weyl multiplet~\cite{Rocek:1985bk}.
The vector compensator~$\mathcal{V}_\text{c}$ in Section~\ref{sec:counterterm} decomposes into a vector multiplet and a chiral multiplet~\cite{Erdmenger:2002ex}.
The auxiliary fields~$Y_{ij}$ and~$Y^{ij}$ in~(\ref{Vc1}) and~(\ref{Vc2}) violate the physical reality condition and hence violate unitarity (as $H$ does in 3$d$).
They descend to an auxiliary field in the 3$d$ vector multiplet that violates the physical reality condition.
It seems likely that the off-shell Poincar\'e supergravity (or at least its supersymmetric background) considered in~\cite{Closset:2012vg} can be obtained from 3$d$ conformal supergravity with the 3$d$ vector multiplet as a compensator.
We conjecture that the imaginary part of $\log Z^\mathcal{I}_\text{SUSY}[\BS^4]$ arises from a counterterm that corresponds to the $Z$-$Z$ Chern-Simons term.

\subsection{Dependence of the SUSY interface partition function on the chirality condition}
\label{sec:dependence}

The full 4$d$ $\mathcal{N}=2$ superconformal algebra is $SU(2,2|2)$.%
\footnote{%
We do not distinguish between a group and its Lie algebra, and ignore the global structure of the former.
}
The Janus interface of our interest preserves the 3$d$ $\mathcal{N}=2$ superconformal algebra $OSp(2|4)_{sc}$.
The massive subalgebra~$OSp(2|4)_m$
of the $SU(2,2|2)$ is generated by SUSY parameters given by~(\ref{epsilon-massive}) and~(\ref{chi-Killing}).
A chirality condition, (\ref{eq:chi-zero-chirality}) or (\ref{eq:chi-zero-chirality-alternative}), further restricts the symmetry to~$OSp(2|2)_m$.%
\footnote{%
The algebra~$OSp(2|2)_m$ coincides with the intersection of $OSp(2|4)_m$ and~$OSp(2|4)_{sc}$.
}

The localization result~(\ref{eq:partition-function-janus})
$$
Z_\text{SUSY}^\mathcal{I} [\BS^4]
=
e^{K(\tau_+, \bar{\tau}{}_-)/12}
$$
for the SUSY interface partition function was obtained by imposing the chirality condition~(\ref{eq:chi-zero-chirality}),
$P_L \chi_0=0$,
 on the SUSY parameter.
We point out that if we instead impose the alternative condition~(\ref{eq:chi-zero-chirality-alternative}), $P_R \chi_0=0$, we
obtain
$$
Z_\text{SUSY}^\mathcal{I}[\BS^4]
=
e^{K(\tau_-, \bar{\tau}{}_+)/12} \ ,
$$
which means that the roles of the north and south poles get exchanged.
Since $K(\tau_+, \bar{\tau}{}_-)^*= K(\tau_-, \bar{\tau}{}_+)$, the phase of the supersymmetric partition function depends on the choice of the chirality condition, or equivalently the choice of~$OSp(2|2)_m$.

In the absence of an interface, the role of a chirality condition is to choose the point~$x^{\mu}=0$ and its antipodal point as the special points to which various quantities such as the on-shell action and the instanton partition functions ``localize''.
Once the condition is imposed, the SUSY parameters generate an~$OSp(2|2)_m$ subalgebra of the massive subalgebra~$OSp(2|4)_m$.
The bosonic factor $Sp(2)\simeq SO(3)$ contains the isometries that preserve the two special points.
If we do not impose either the condition $P_L \chi_0=0$ or $P_R \chi_0=0$ we obtain, in the absence of an interface, the same partition function; indeed given a non-zero $\chi_0^j$ we can take, as the special point (the north pole), the solution $x^{\mu}$ to the equation 
\begin{align} \label{eq:chi-L-R}
\left( \epsilon^j\propto P_L \chi^j \propto\right)\quad P_L \chi_{0}^j + \frac{\mathrm{i}}{2r} x_{\mu} \Gamma^{\mu} P_R \chi_{0}^j =0 \ .
\end{align}
This is a system of four equations ($j=1,2$ and two components for a chiral spinor) for four unknowns $x^{\mu}$ ($\mu=1,\ldots,4$) and (at least generically) has a solution.

\subsection{Conformal anomaly in the presence of an interface}\label{sec:anomaly-interface}
In Section~\ref{sec:IE-sphere} we derived the relation \eqref{eq:relation-ie-partition-function-nonSUSY} between the interface entropy and the sphere partition functions on $\BS^4$ using the dimensional regularization.
\eqref{eq:relation-ie-partition-function-nonSUSY} provides us an easier and more pragmatic way to calculate the interface entropy than the original definition \eqref{ICFT_RE-limit}, and is the key to proving the equivalence between the interface entropy and Calabi's diastasis in this paper.
The crucial point of the derivation in \cite{Kobayashi:2018lil} is that in the dimensional regularization there are no conformal anomalies, hence one can ignore a possible contribution from the conformal anomaly in calculating the interface entropy.
The anomaly is automatically incorporated as poles at even dimensions in the final result.
The validity of the approach in \cite{Kobayashi:2018lil} was supported by the holographic computation, so we believe \eqref{eq:relation-ie-partition-function} universally holds in any dimensions.
In our case the holographic calculation of the sphere partition function and the interface entropy in Section~\ref{sec:holography} gives an additional evidence for the relation \eqref{eq:relation-ie-partition-function}.

On the other hand, the use of the dimensional regularization in Section~\ref{sec:IE-sphere} obscures how conformal anomalies could have appeared if the same line of argument would be followed in four dimensions.
So it would be instructive to revisit the derivation in Section~\ref{sec:IE-sphere}, but now in $d=4$ dimensions.

First the partition function is no longer invariant under the CHM map and gets a contribution from the anomaly:
\begin{align} 
	Z^\text{(ICFT)}[\CM_n] = Z^\text{(ICFT)}[\BS^4_n] \times e^{-\int_{\BS^4_n}\d^4 x\,\mathcal{A}[g_{\mu\nu}]}\ .
\end{align}
The conformal anomaly is a functional of the background metric $\CA[g_{\mu\nu}]$.
In CFT without an interface, it transforms under an infinitesimal conformal transformation $\delta g_{\mu\nu} = 2\sigma\,g_{\mu\nu}$ as
\begin{align}
	\frac{\delta \CA^\text{(CFT)}}{\delta \sigma} = a\, E + c\, I \ ,
\end{align}
where $a$ and $c$ are the central charges, and $E$ and $I$ are the Euler density and Weyl invariant in four dimensions \cite{Deser:1993yx}.
In ICFT, there is an additional contribution localized on an interface to the conformal anomaly
\begin{align}
	\CA = \CA^\text{(CFT)} + \delta_\CI\,\CA^{(\CI)} \ ,
\end{align}
where $\delta_\CI$ is the delta function supported on the interface.
The anomaly gives rise to an additional contribution to the entanglement entropy:
\begin{align}
	S_E^\text{(ICFT)} = \cdots - \lim_{n\to 1}\,\frac{1}{1-n}\,\left[ \left(\int_{\BS_n^4}\d^4 x - n\int_{\BS^4}\d^4 x\right)\,\CA \right] \ ,
\end{align}
whose ambient part $\CA^\text{(CFT)}$ are shown to yield the logarithmically UV divergent term \cite{Solodukhin:2008dh}, but it is cancelled by the same anomaly from CFT$_\pm$ in the interface entropy \eqref{Def_IE}.
The localized term $\CA^{(\CI)} $, on the other hand, remains unsubtracted and contributes to $S_\CI$.

The conformal anomaly also modifies the transformation law of the one-point function~$ \langle\, T^{\mu\nu}\,\rangle_{\BS^4}^\text{(ICFT)}$ from \eqref{VEV_stress_tensor},
\begin{align}\label{VEV_stress_tensor_anom}
	\langle\, T_{\mu\nu}\,\rangle^{(\text{ICFT})}_{\BS^4} =(\text{Weyl factor})^2 \langle\, T_{\mu\nu}\,\rangle^{(\text{ICFT})}_{\BR^4} + \CA_{\mu\nu}|_{\BS^4} = \CA_{\mu\nu}|_{\BS^4} \ ,
\end{align}
where $\CA_{\mu\nu}$ is the anomalous part of the stress tensor,
\begin{align}
	\CA_{\mu\nu} \equiv \frac{2}{\sqrt{g}}\, \frac{\delta\, \int \d^4 x\, \CA [g_{\mu\nu}]} {\delta g_{\mu\nu}}\ .
\end{align}
It also consists of the ambient and localized terms:
\begin{align}
	\CA_{\mu\nu} = \CA_{\mu\nu}^{(\text{CFT})} + \delta_\CI\,\CA_{\mu\nu}^{(\CI)} \ .
\end{align}
The explicit form of its ambient part can be found in \cite{Brown:1977sj,Herzog:2013ed}.
On $\BS^4$ the ambient part $\CA_{\mu\nu}^{(\text{CFT})}$ can be fixed from the type-$A$ trace anomaly~\cite{Brown:1977sj,Herzog:2013ed} as
\begin{equation}
 \CA_{\mu\nu}^{(\text{CFT})}  = - \frac{a}{(4\pi)^2}\left[
 g_{\mu\nu}\left(\frac{R^2}{2}- R^{\rho\lambda}R_{\rho\lambda}\right)  + 2 R_{\mu\lambda}R_\nu{}^\lambda - \frac43 R\, R_{\mu\nu} 
 \right] \ .
\end{equation}
On the other hand the localized anomaly $\CA_{\mu\nu}^{(\CI)}$ associated with the interface is not known except for the trace part in BCFT
\begin{equation}\label{ACI-trace}
\CA^{(\CI)}{}^\mu{}_{\mu} = \frac{1}{16\pi^2} \left ( a\, E_4^\text{(bry)} - b_1\,  {\rm tr} \,\hat K^3 - b_2\, h^{\alpha\gamma} \hat K^{\beta\delta} W_{\alpha\beta\gamma\delta}  \right) \ .
\end{equation}
We refer to~\cite{Herzog:2017xha} for the definitions of various symbols.
See also the paper~\cite{Herzog:2020wlo} that focuses on interfaces.
The quantity $\CA_{\mu\nu}^{(\CI)}$ should be a geometric functional of the background metric and the extrinsic curvature, but it remains open how to fix the explicit form.

A moment's thought shows that the ambient terms $\CA_{\mu\nu}^{(\text{CFT})}$ are there both in ICFT and CFT$_\pm$ with the same value, hence cancel out in the interface entropy \eqref{Def_IE} in the same way as $\CA^\text{(CFT)}$ in the previous paragraph.

Collecting the possible contributions from the localized anomalous term, we find a deviation $\Delta S_\CI$  from \eqref{eq:relation-ie-partition-function-nonSUSY}:
\begin{align}
	\begin{aligned}
		\Delta S_\CI &= \int_{\BS^4} \d^4x\,\delta(\phi - \pi/2)\,\sin^2\theta\,\CA^{(\CI)\,\tau\tau} \\
			&\qquad - \lim_{n\to 1}\,\frac{1}{1-n}\,\left[ \left(\int_{\BS_n^4}\d^4 x - n\int_{\BS^4}\d^4 x\right)\,\delta(\phi - \pi/2)\,\CA^{(\CI)} \right] \ .
	\end{aligned}
\end{align}
Compared with the dimensional regularization result, this result indicates that the anomalous terms from the interface-localized anomaly $\CA^{(\CI)}$ should integrate to zero on a sphere while the ambient anomalous parts nicely cancel out in the definition of $S_\CI$.

It would be nice to determine the explicit forms of $\CA^{(\CI)}$ and $\CA_{\mu\nu}^{(\CI)}$ from~(\ref{ACI-trace}) along the lines of \cite{Brown:1977sj,Herzog:2013ed} and directly check that it does not contribute to the interface entropy.

\acknowledgments
We would like to thank Y.\,Kazama, K.\,Maruyoshi, Y.\,Nakayama and I.\,Yaakov for valuable discussions.
We thank C.\,Bachas for useful communication.
The work of T.\,O.  is supported in part by the JSPS Grant-in-Aid for Scientific Research (C) No.JP16K05312.
The work of T.\,N. is supported in part by the JSPS Grant-in-Aid for Scientific Research (C) No.19K03863 and the JSPS Grant-in-Aid for Scientific Research (A) No.16H02182.
We thank the Yukawa Institute for Theoretical Physics at Kyoto University,
where a part of this work was done during the workshop YITP-T-19-03 ``Quantum Information and String Theory 2019."
We also thank the participants of the conference ``Strings and Fields 2019'' for stimulating discussions.


\appendix

\section{Supersymmetry and supergravity} \label{app:SUSY-SUGRA}
\subsection{Notations and conventions}\label{app:notations}
We use the notation and the convention in~\cite{VanProeyen-note, Freedman:2012zz} unless otherwise noted.
Complex conjugation is indicated by~$*$ and hermitian conjugation by~$\dagger$.
The imaginary unit is ${\rm i}$.
Coordinates have indices $\mu, \nu,\ldots$.
The vielbein is $e_\mu{}^a$, and its inverse is $e_a{}^\mu$ with tangent (or flat) space indices $a,b,\ldots$.

\subsubsection{Gamma matrices}

In Minkowski signature we have $\eta^{ab}=\text{diag}(-1,1,1,1)$ with $a,b=0,\ldots,3$, while in Euclidean signature~$\eta^{ab}=\text{diag}(1,1,1,1)$ with $a,b=1,\ldots,4$.
The gamma matrices~$\gamma^\mu$ (with a Greek alphabet) satisfy 
\begin{equation}
\{\gamma^{\mu},\gamma^{\nu}\}=2g^{\mu\nu} \ ,
\end{equation}
while the gamma matrices $\gamma^a$  (with a Latin alphabet) satisfy%
\footnote{%
In the Weyl representation we have
\begin{align} \label{gamma-weyl}
\gamma^{a}=
\begin{pmatrix}
0 & \s^{\mu}
\\
\bar{\s}^{\mu} & 0
\end{pmatrix},
\end{align}
where $\sigma^{\mu}=({\sigma}^{1},{\sigma}^{2},{\sigma}^{3},\i), \bbar{\sigma}^{a}=({\sigma}^{1},{\sigma}^{2},{\sigma}^{3},-\i)$,
and $\s^{i}\, (i=1,2,3)$ are Pauli matrices.
}
\begin{equation}\label{Clifford-flat}
\{\gamma^{a},\gamma^{b}\}=2 \eta^{ab}   \ .
\end{equation}
They are related as
\begin{equation}
\gamma^\mu = \gamma^a e_a{}^\mu \ .
\end{equation}
In flat space there is no distinction.
The matrix $\gamma^a$ is anti-hermitian if~$a=0$, and is hermitian otherwise.
We have $\gamma^{a=0}=-{\rm i}\gamma^{a=4}$.
In terms of the chirality matrix $\gamma_{*}={\rm i} \gamma_{0}\gamma_{1}\gamma_{2}\gamma_{3}=\gamma_{1}\gamma_{2}\gamma_{3}\gamma_{4}$%
\footnote{%
Here each gamma matrix is $\gamma_a$.
More generally $\gamma_\mu$ and $\gamma_a$ should be distinguished based on the context.
As in~\cite{Gomis:2012wy}, we sometimes write $\Gamma^a$ for $\gamma^a$.
}
 we define the chirality projections $P_{L}, P_{R}$ by
\begin{align}
P_{L}=\frac{1}{2}\,(1+\gamma_{*})\ ,
\qquad
P_{R}=\frac{1}{2}\,(1-\gamma_{*})\ .
\end{align}

\subsubsection{$SU(2)_{R}$ multiplets}
We denote by $i,j,\dots$ $SU(2)_{R}$ doublet indices.
We regard an $SU(2)_{R}$ triplet as a three-component vector, from which we can form a tensor with two indices
\begin{align}
{Y_{i}}^{j}={{\vec{\tau}}_{i}}{}^{j}\cdot\vec{Y}\ ,
\end{align}
where ${{\vec{\tau}}_{i}}{}^{j}=\i\,{{\vec{\s}}_{i}}{}^{j}$.
Let ${\ve}^{ij}$ and ${\ve}_{ij}$ be anti-symmetric tensors such that
\begin{align}
{\ve}^{12}={\ve}_{12}=1\ .
\end{align}
Sometimes but not always, we use them to raise and lower doublet indices, as in
\begin{align}
\vec{\tau}^{ij}={\ve}^{ik}{{\vec{\tau}}_{k}}{}^{j}
=(\vec{\tau}_{ij})^{*} = \varepsilon^{ik}\varepsilon^{jl} \vec\tau_{kl}\ .
\end{align}
Using $\vec{\tau}^{\,ij}$ we can convert the $SU(2)_{R}$ triplets into symmetric matrices
\begin{align}
Y^{ij}=\vec{\tau}{}^{\,ij}\cdot\vec{Y} \ .
\end{align}
We note a useful formula
\begin{align}
A_{ij}B^{jk}=\delta_{i}^{k}\vec{A}\cdot\vec{B}
+
(\vec{A}\times\vec{B})\cdot{\vec{\tau}_{i}}{}^{k}\ .
\end{align}

\subsubsection{Conjugations in Minkowski signature}
The charge conjugation matrix~$C$ satisfies%
\footnote{%
We choose $t_0=1$, $t_1=-1$, etc. in Table~3.1 of~\cite{Freedman:2012zz}.
}
\begin{align}
C C^{\dag}=1 \,
\quad
C^{T}=-C\ ,
\quad
C \gamma_{\mu}C^{-1} 
=-\gamma_\mu^T\ .
\end{align}
We also introduce%
\footnote{%
In the Weyl representation~(\ref{gamma-weyl}), we can take $C={\rm i}\gamma^3\gamma^1$, $B=(\gamma^{0}\gamma^{1}\gamma^{3})^{-1}$.
}
\begin{align}
B=\i\,C\gamma^{0}\ .
\end{align}
In Minkowski signature we define the charge conjugation $\Psi^{C}$ of a 4-component spinor $\Psi$ by
\begin{align}
\Psi^{C}=B^{-1}\Psi^{*}\ .
\end{align}
We have~$(\Psi^C)^C=\Psi$, $(\gamma_{\mu_1}\ldots\gamma_{\mu_N}\Psi)^C = \gamma_{\mu_1}\ldots\gamma_{\mu_N}\Psi^C$.
The matrix~$B$ satisfies the relation
\begin{align}
B^{-1}(\gamma^{\mu})^{*}B=\gamma^{\mu}\ .
\end{align}
We indicate the Weyl conjugate of a spinor by a bar:
\begin{equation}\label{Weyl-conj-def-Psi}
\bar{\Psi} :=\Psi^{T}C\ .
\end{equation}
For two spinors $\epsilon$ and $\eta$, we have
\begin{equation} \label{bilinear-conjugate}
(\bar{\epsilon}\gamma_{\mu_1}\ldots\gamma_{\mu_N}\eta)^* = \pm \overline{\epsilon^C} \gamma_{\mu_1}\ldots\gamma_{\mu_N}\eta^C \ ,
\end{equation}
where we take the upper sign when they are both odd and the lower sign otherwise.

\subsection{Supersymmetry parameters}

In Minkowski signature the parameters for Poincar\'e supersymmetry satisfy
\begin{align}
(\e^{i})^{C}=\e_{i}\ .
\end{align}
For such parameters, the Weyl conjugate~(\ref{Weyl-conj-def-Psi}) coincides with the Dirac conjugate:
\begin{equation}
\bar{\e}{}^{i}=(\e_{i})^{\dag}\i\gamma^{0}  \ .
\end{equation}
The parameters for special superconformal symmetry similarly satisfy
\begin{align}
(\eta^{i})^{C}=\eta_{i}\ .
\end{align}

Both in Minkowski and Euclidean signatures, these parameters are chiral:
\begin{equation}
\epsilon^i = P_L \epsilon^i \ , \quad
\epsilon_i = P_R \epsilon_i \ , \quad
\eta^i = P_R \eta^i \ , \quad
\eta_{i} = P_L \eta_{i} \ .
\end{equation}

\subsection{$\mathcal{N}=2$ supermultiplets} \label{sec:N=2mult}
In the rest of Appendix~\ref{app:SUSY-SUGRA}, we assume that the background values of the Weyl multiplet are all zero except the metric and the vielbein.
We now explain $\mathcal{N}=2$ vector and chiral multiplets following~\cite{VanProeyen-note}.
Formulas are given for the Minkowski signature and for anti-commuting parameters satisfying
$\eta^i=\frac14 \gamma^\mu\nabla_\mu\epsilon^i$, $\eta_i=\frac14 \gamma^\mu\nabla_\mu\epsilon_i$~\cite{Gomis:2014woa}.
Care must be taken when applying them in Euclidean signature and with commuting SUSY parameters.
The transformations valid in these cases are obtained from the formulas in~\cite{VanProeyen-note} by explicitly computing ``h.c." by~(\ref{bilinear-conjugate}) to have expressions with odd parameters on the left.  For example, the ``h.c.'' of $\bar{\epsilon}_{i} \gamma_{\mu}\Omega_{j}$ with $\epsilon_i$ and $\Omega_j$ odd gives the expression $\bar{\epsilon}^{i} \gamma_{\mu}\Omega^{j}$, which is valid in Euclidean signature and with $\epsilon_i$ even.

\subsubsection{Vector multiplet}\label{subsec:vector-multiplet}
A vector multiplet has $(X,\Omega_{i},A_{\mu},Y_{ij})$ as its components.
The spinor~$\Omega_{i}$ is the left-handed gaugino, and its charge conjugate~$\Omega^{i}$ is right-handed.
We use hermitian generators $T_I$ such that $[T_I, T_J]={\rm i} f_{IJ}{}^K T_K$ and expand $X=T_I X^I$, $A_\mu = T_I A^I_\mu$, etc.%
\footnote{%
Our hermitian generators~$T_I$ are related to the anti-hermitian generators~$t_I$ in~\cite{VanProeyen-note,Freedman:2012zz} as $t_I= -\i T_I$.
Most of the formulas in the references are given in terms of the coefficient fields~$X^I$, $A_\mu^I$, etc.
}
Their SUSY transformations are~\cite{VanProeyen-note}
\begin{align}
\delta X^{I}&=\frac{1}{2}\,\bbar{\epsilon}^{i}\Omega_{i}^{I},
\label{eq:susy-transf-x-comp2}
\\
\delta\Omega_{i}^{I}
&=\sla{D}X^{I}\epsilon_{i}
+\frac{1}{4}\,\gamma^{\mu\nu}F_{\mu\nu}^{I}\,\varepsilon_{ij}\epsilon^{j}
+\frac{1}{2}\,Y_{ij}^{I}\,\epsilon^{j}
+X^{J}\bbar{X}^{K}{f_{JK}}^{I}\varepsilon_{ij}\,\epsilon^{j}
+2 X^I \eta_i
\ ,
\\
\dlt A_{\mu}^{I}&=\frac{1}{2}\,\varepsilon^{ij}\,\bar{\epsilon}_{i} \gamma_{\mu}\Omega_{j}^{I}
+ \frac{1}{2}\,\varepsilon_{ij}\,\bar{\epsilon}{}^{i} \gamma_{\mu}\Omega^{jI}
\ ,
\\
\delta\vec{Y}^{I}
&=
\frac{1}{ 2}\vec{\tau}{}^{\, ij}\,\bbar{\e}_{i} \sla{D}\Omega_{j}^{I}
-
{f_{JK}}^{I}\,\vec{\tau}_i{}^{j}\,\bbar{\e}_{j}\,X^{J}\Omega^{iK}
+\text{h.c.}
\ .
\label{eq:susy-transf-y-comp2}
\end{align}
In Minkowski space we have $(\Omega_i^I)^C=\Omega^{Ii}$.

\subsubsection{Chiral multiplet}\label{subsec:chiral-multiplet}
A chiral multiplet~$\mathcal{A}$ has $(A,\Psi_{ i},B_{ij},F^{ -}_{ab},\Lambda_{i}, C)$ as its components.
Their SUSY transformations are~\cite{Breitenlohner:1980ej,VanProeyen-note}
\begin{align}
	\delta A&=\frac{1}{2}\,\bar{\e}^{i}\Psi_{i}\ ,\\
	\delta\Psi_{i}&=\sla{\nabla}(A\,\e_{i})+\frac{1}{2}\,B_{ij}\,\e^{j} +\frac{1}{4}\,\Gamma^{ab}F_{ab}^{-}\,\ve_{ij}\,\e^{j} +(2w-4)\,A\,\eta_{i}\ ,
\label{eq:susy-variation-psi} \\
	\delta B_{ij}&=\bar{\e}_{(i}\,\sla{\nabla}\,\Psi_{j)}-\bar{\e}^{k}\,\Lambda_{(i}\,\ve_{j)k}+2(1-w)\,\bar{\eta}_{(i}\,\Psi_{j)}\ , \\
	\delta F_{ab}^{-}&=\frac{1}{4}\,\ve^{ij}\,\bar{\e}_{i}\,\sla{\nabla}\,\Gamma_{ab}\,\Psi_{j}
		+\frac{1}{4}\,\bar{\e}^{i}\,\Gamma_{ab}\Lambda_{i} - \frac{1}{2}\,(1+w)\,\ve^{ij}\,\bar{\eta}_{i}\,\Gamma_{ab}\Psi_{j}\ , \\
	\delta\Lambda_{i}&= -\frac{1}{4}\,\Gamma^{ab}\,\sla{\nabla}(F_{ab}^{-}\,\e_{i}) - \frac{1}{2}\,\sla{\nabla}\,B_{ij}\,\ve^{jk}\,\e_{k} + \frac{1}{2}\,C\ve_{ij}\,\e^{j} \notag \\
		&\qquad -(1+w)\,B_{ij}\,\ve^{jk}\,\eta_{k} +\frac{1}{2}\,(3-w)\,\Gamma^{ab}F_{ab}^{-}\,\eta_{i}\ ,
	\label{eq:susy-variation-lambda} \\
	\delta C&= -\nabla_{\mu}(\ve^{ij}\,\bar{\e}_{i}\,\gamma^{m}\Lambda_{j}) +(2w-4)\,\ve^{ij}\,\bar{\eta}_{i}\,\Lambda_{j}\ ,
\end{align}
where $w$ is the Weyl weight of the multiplet.

An anti-chiral multiplet~$\overline{\mathcal{A}}$ has $(\overline{A},\Psi^i,B^{ij},F^{ +}_{ab},\Lambda^i, \overline{C})$ as its components.
In Minkowski signature, its transformations are obtained from those of the chiral multiplet~$\mathcal{A}$ by complex or charge conjugation.
In Euclidean signature, the transformations are obtained from those in Minkowski signature by the procedure described at the beginning of this subsection.

\subsection{Tensor calculus for chiral multiplets}\label{sec:tensor-calc}
Given two chiral multiplets $\mathcal{
A}$ and $\mathcal{B}$ with vanishing fermionic components
\begin{align}
\mathcal{A}&=(A|_{\mathcal{A}},\,\Psi_{i}|_{\mathcal{A}}=0,\,B_{ij}|_{\mathcal{A}},\, F^{-}_{ab}|_{\mathcal{A}},\, \Lambda_{i}|_{\mathcal{A}}=0,\, C|_{\mathcal{A}})\ ,
\\
\mathcal{B}&=(A|_{\mathcal{B}},\, \Psi_{i}|_{\mathcal{B}}=0,\, B_{ij}|_{\mathcal{B}},\, F^{-}_{ab}|_{\mathcal{B}},\, \Lambda_{i}|_{\mathcal{B}}=0,\,  C|_{\mathcal{B}})\ ,
\end{align}
the product chiral multiplet $\mathcal{A}\mathcal{B}$ is given as~\cite{deRoo:1980mm}
\begin{align}
A|_{\mathcal{A}\mathcal{B}}
&=
A|_{\mathcal{A}}\, A|_{\mathcal{B}}\ ,
\\
B_{ij}|_{\mathcal{AB}}&=
A|_{\mathcal{A}}\, B_{ij}|_{\mathcal{B}}
+
A|_{\mathcal{B}}\, B_{ij}|_{\mathcal{A}}\ ,
\\
F^{-}_{ab}|_{\mathcal{AB}}
&=
A|_{\mathcal{A}}\, F^{-}_{ab}|_{\mathcal{B}}
+
A|_{\mathcal{B}}\, F^{-}_{ab}|_{\mathcal{A}}\ ,
\\
C|_{\mathcal{AB}}
&=
A|_{\mathcal{A}}\, C|_{\mathcal{B}}
+
C|_{\mathcal{A}}\, A|_{\mathcal{B}}
-\frac{1}{2}\,\ve^{ik}\ve^{jl}B_{ij}|_{\mathcal{A}}\, B_{kl}|_{\mathcal{B}}
+F^{-}_{ab}|_{\mathcal{A}}\, F^{-ab}|_{\mathcal{B}}\ .
\end{align}
The $n$-th power of a chiral multiplet $\mathcal{A}$~\cite{deWit:1980lyi} is given as
\begin{align}
A|_{\mathcal{A}^{n}}
&=
\left(A|_{\mathcal{A}}\right)^{n}\ ,
\\
B_{ij}|_{\mathcal{A}^{n}}
&=
n\left(A|_{\mathcal{A}}\right)^{n-1}B_{ij}|_{\mathcal{A}}\ ,
\\
F^{-}_{ab}|_{\mathcal{A}^{n}}
&=
n\left(A|_{\mathcal{A}}\right)^{n-1}F_{ab}^{-}|_{\mathcal{A}}\ ,
\\
C|_{\mathcal{A}^{n}}
&=
n\left(A|_{\mathcal{A}}\right)^{n-1}C|_{\mathcal{A}}
-\frac{1}{4}\,n(n-1)\left(A|_{\mathcal{A}}\right)^{n-2}
\left[\ve^{ik}\ve^{jl}B_{ij}|_{\mathcal{A}}B_{kl}|_{\mathcal{A}}-2\left(F_{ab}^{-}|_{\mathcal{A}}\right)^{2}\right]\ .
\end{align}
For fields in the adjoint representation, we should apply these formulas to the coefficients of the generators~$T_I$.

\subsection{Definition of $\mathbb{T}(\log\bar{\Phi})$}
In this appendix we give the expression for~$\mathbb{T}(\log\bar{\Phi})$ computed from an anti-chiral multiplet~$\bar{\Phi}$ with vanishing fermionic and field strength components.
First, the components of $\log\bar{\Phi}$ are given by~\cite{Butter:2013lta}
\begin{align}
\bar{A}|_{\log\bar{\Phi}}&=\log \left(\bar{A}|_{\bar{\Phi}}\right) ,
\\
B^{ij}|_{\log\bar{\Phi}}&=\frac{{B}^{ij}|_{\bar{\Phi}}}{\bar{A}|_{\bar{\Phi}}} ,
\\
\bar{C}|_{\log\bar{\Phi}}&=
\frac{\bar{C}|_{\bar{\Phi}}}{\bar{A}|_{\bar\Phi}}
+
\frac{1}{4\left(\bar{A}|_{\bar{\Phi}}\right)^{2}}
\varepsilon_{ik}\varepsilon_{jl}\left(B^{ij}|_{\bar{\Phi}}\right)\left(B^{kl}|_{\bar{\Phi}}\right) .
\end{align}
The chiral multiplet $\mathbb{T}(\text{anti-chiral multiplet})$ is the so-called $\mathcal{N}=2$ kinetic multiplet~\cite{deWit:1980lyi}.
The components of the kinetic multiplet made from $\log\bar{\Phi}$ are given as~\cite{Butter:2013lta}
\begin{align}
A|_{\mathbb{T}(\log\bar{\Phi})}
&=
\bar{C}|_{\log\bar{\Phi}} \ ,
\\
B_{ij}|_{\mathbb{T}(\log\bar{\Phi})}
&=
-2\ve_{ik}\ve_{jl}\dal_{C}B^{kl}|_{\log\bar{\Phi}} \ ,
\\
C|_{\mathbb{T}(\log\bar{\Phi})}
&=
4\dal_{C}\dal_{C}\bar{A}|_{\log\bar{\Phi}} \ ,
\end{align}
where $\dal_{C}$ is the so-called conformal d'Alembertian.

\section{Conformal transformations between $\BS^4$ and the flat space}\label{app:conf}

Let us consider the embedding coordinates $Y^{M}$ ($M=1,\ldots,5$) for $\BS^4$ satisfying
\begin{equation}
\sum (Y^M)^2 = r^2 \ ,
\qquad
\d s^2_{\BS^4} = \sum (\d Y^M)^2 \ .
\end{equation}
Recall the coordinates $x^\mu$ used in Section~\ref{sec:massive} and  $y^\mu$ used in Section~\ref{sec:off-shell-flat}.
We define $x=(\sum_\mu (x^\mu)^2)^{1/2}$,  $y=(\sum_\mu (y^\mu)^2)^{1/2}$.
We also define two functions of a single variable $z$:
\begin{equation}
f(z):= \frac{1}{1+\frac{z^2}{4r^2}} \ , \qquad
g(z):=r\, \frac{1- \frac{z^2}{4r^2}}{1+ \frac{z^2}{4r^2}}\ .
\end{equation}
By~(\ref{x-theta}) we have
\begin{equation}
f(x)= \cos^2\frac{\theta}{2} \ .
\end{equation}
The coordinates $x^\mu$ and $y^\mu$ are related to $Y^M$ as
\begin{equation}
\begin{pmatrix}
Y^1 \\
Y^2 \\
Y^3 \\
Y^4 \\
Y^5 \\
\end{pmatrix}
=
\begin{pmatrix}
f(x) 
\begin{pmatrix}
x^1 \\
x^2 \\
x^3 \\
x^4 \\
\end{pmatrix}
\\
  g(x)
\end{pmatrix}
=
\begin{pmatrix}
   g(y)
\\
f(y) 
\begin{pmatrix}
y^4 \\
y^1 \\
y^2 \\
y^3 \\
\end{pmatrix}
\end{pmatrix} \,.
\end{equation}
From the relations $f(x)x^1=g(y)$ and $g(x)=f(y)y^3$ we find
\begin{equation}
y^3 =\frac{2r \cos\theta}{1+\sin\theta\cos\theta_1} \ , \qquad 
f(y)= \frac{1}{2}(1+ \sin\theta\cos\theta_1) \ ,
\end{equation}
where $\cos\theta_1=x^1/x $.
Then
\begin{equation} \label{delta-theta-y3}
\frac{1}{r} \delta\left(\theta-\frac\pi 2\right) =
 \frac{1}{f(y)}\,
 \delta(y^3) \ ,
 \qquad
 \frac{1}{r^2} \delta'\left(\theta-\frac\pi 2\right) =
- \frac{1}{f(y)^2}\,
 \delta'(y^3) \ .
\end{equation}
Since the sphere metric can be written as
$
\d s^2_{\BS^4} = f(y)^2 \d y^\mu \d y^\mu
$,
$f(y)$ is the conformal factor that relates the metrics in Sections~\ref{sec:off-shell-flat} and~\ref{sec:massive}.
The Weyl weights of $B_{ij}^{(\tau)}$, $B^{(\tau)ij}$, $C^{(\tau)}$, and $\overline{C}{}^{(\tau)}$, are $1,1,2,2$, respectively~\cite{VanProeyen-note}.
The identities~(\ref{delta-theta-y3}) then imply that~(\ref{BC-flat-delta}) and~(\ref{BC-sphere-delta}) are related by the Weyl transformation.

\section{Details on the supersymmetric R\'enyi entropy}\label{sec:SRE-details}


In this appendix we provide some details that we use in Section~\ref{sec:SRE} when we discuss the supersymmetric R\'enyi entropy.

\subsection{SUSY background on the branched 4-sphere}\label{sec:SRE-background}
To complete the definition of the supersymmetric R\'enyi entropy~(\ref{SRE-absolute-value}), we review the relevant part of the supersymmetric background~$\widetilde{\BS}^4_n$ that regularizes the $n$-fold branched cover of the 4-sphere with metric~(\ref{Branched_Sphere}).
For simplicity we set the radius of the sphere to one.
First let us consider a four-manifold $X_4$ that is a torus fibration over a 2$d$ surface.
One can pick coordinates $(\phi_1, \phi_2)$ and $(\eta, \rho)$ for the torus and the surface respectively, and introduce the metric of the form~\cite{Pestun:2014mja}
\begin{align}\label{X4}
	\begin{aligned}
\d s^2_{X_4} &= \sin^2\rho\, \left(\epsilon_1^{-2}\, \cos^2\eta\,\d\phi_1^2 + \epsilon_2^{-2}\, \sin^2\eta\, \d\phi_2^2\right) \\
	&\qquad \qquad + \left(f_1(\eta, \rho) \sin\rho\,\d\eta + f_3(\eta, \rho)\,\d\rho \right)^2 + f_2(\eta, \rho)^2\, \d \rho^2 \ ,
	\end{aligned}
\end{align}
where $\epsilon_1, \epsilon_2$ are constants and $f_1, f_2, f_3$ are functions on the surface.

We regularize the singularity of the branched sphere metric~\eqref{Branched_Sphere} in four dimensions by replacing it with the resolved branched sphere $\widetilde \BS^n_4$,
\begin{align}\label{tS4n-metric}
	\d s^2_{\widetilde \BS^n_4} = f(\theta)^2\,\d \theta^2 + n^2\sin^2\theta\, \d \tau^2 + \cos^2\theta\,\left(\d \phi^2 + \sin^2\phi\,\d \chi^2 \right) \ ,
\end{align}
where we introduced a smooth function $f(\theta)$ such that
\begin{align}
	f(\theta \to 0) = n \ , \qquad f(\theta \gg \delta ) = 1 \ ,
\end{align}
for a small parameter $\delta \ll 1$.
By changing the coordinates via
\begin{align}
	\sin\theta = \sin\eta\, \sin\rho \ , \qquad \tan \phi = \cos\eta\, \tan \rho \ ,\qquad \chi = \phi_1 \ , \qquad \tau = \phi_2 \ ,
\end{align}
the metric takes the form~\eqref{X4} with $\epsilon_1 = 1, \epsilon_2 = 1/n$ and%
\footnote{%
The expressions in~(\ref{eq:f123}) are equivalent to (C.3) of~\cite{Huang:2014pda} with~$(f_1,f_2,f_3)=(F,G,H)_\text{there}$.
}
\begin{align}\label{eq:f123}
	\begin{aligned}
		f_1(\eta, \rho) &= \left( \frac{f(\theta)^2\,\cos^2\eta + \sin^2\eta\,\cos^2\rho}{\cos^2\eta + \sin^2\eta\,\cos^2\rho} \right)^{1/2} \ , \\
		f_2(\eta, \rho) &= \frac{f(\theta)}{f_1(\eta, \rho)} \ ,\\
		f_3(\eta, \rho) &= \frac{(f(\theta)^2 - 1)\,\sin(2\eta)\,\cos\rho}{2 f_1(\eta, \rho)\, (\cos^2\eta + \sin^2\eta\,\cos^2\rho)} \ .
	\end{aligned}
\end{align}
For $n=1$ and $f(\theta)=1$, (\ref{tS4n-metric}) reduces to the round sphere metric~(\ref{eq:S4-metric}).
We note that the interface is placed at $\phi=\pi/2$ or equivalently at $\rho=\pi/2$.

Part of supersymmetries can be preserved by tuning on the background supergravity fields $V_\mu^{ij}, A_\mu, T^\pm_{\mu\nu}$ and $D$ in the Weyl multiplet~\cite{Pestun:2014mja}.%
\footnote{%
In the singular limit $\delta\to 0$ the SUSY background of $\widetilde \BS^n_4$~\cite{Huang:2014pda} reduces, away from the singularities, to
$	(V_\tau)^i_{~j} = A_\tau \text{diag}(1,-1)$, 
$ A_\tau = \frac{n-1}{2}$, and $ T^\pm_{\mu\nu} = D = 0  $.
}

\subsection{Vanishing of one-point functions}\label{ss:supercurrent}

We now show that the second term in~(\ref{ICFT_SRE-derivative}) vanishes.

An $\CN=2$ SCFT has the supercurrent multiplet~\cite{Sohnius:1978pk,Fisher:1982fu,Kuzenko:1999pi,Antoniadis:2010nj}%
\footnote{%
It was shown in~\cite{Gomis:2014woa} that the system with the massive superalgebra~$OSp(2|4)_m$ symmetry corresponds to an off-shell formulation of $\mathcal{N}=2$ Poincar\'e supergravity with vector and tensor multiplets used as compensators.
The SUSY parameters for~$OSp(2|4)_m$ are compatible with this off-shell formulation, but not with the off-shell formulation that involves a non-linear multiplet or a hypermultiplet as a compensator.
For a similar hidden dependence of theory on the off-shell formulation of supergravity, see~\cite{Lambert:2005dx}.
The compensating tensor multiplet in general affects the conservation equation for the supercurrent~\cite{Butter:2010sc}.
In our set-up, however, the tensor multiplet does not actually couple to the field theory and hence does not affect the conservation equation.
}
\begin{align}\label{N=2_Supercurrent}
	\CJ = (T_{\mu\nu},\, S_{\mu}^i,\, j_{\mu}^{ij},\, j_\mu,\, J,\, j^{i},\, j^\pm_{\mu\nu} ) \ ,
\end{align}
which includes the stress tensor $T_{\mu\nu}$, the supersymmetry current $S_\mu^i$, the $SU(2)_R$ current~$j_{\mu}^{ij}$, and the $U(1)_R$ current $j_\mu$. 
The supercurrent also contains a real scalar $J$, self-dual and anti-self-dual anti-symmetric tensors $j^\pm_{\mu\nu} $, and a spinor $j^i$.
The spinorial operators are chiral: $S^i_\mu = P_R S^i_\mu$, $j^i = P_R j^i$.
We suppressed their conjugates~$S_{\mu i} = P_L S_{\mu i}$ and $j_i = P_L j_i$ in~(\ref{N=2_Supercurrent}).

The supercurrent multiplet couples to the Weyl multiplet.
Among the fields in the Weyl multiplet, those which couple to the SCFT supercurrent are
\begin{align}\label{Weyl-multiplet-linearized}
	\CW = (g_{\mu\nu},\, \psi_{\mu}{}^i,\, V_{\mu\,i}{}^{j},\, A_\mu,\, D,\, \chi^i,\, T^\pm_{\mu\nu}) \ .
\end{align}
Spinorial fields are chiral: $\psi_{\mu}{}^i = P_L \psi_{\mu}{}^i$, $\chi^i=P_L\chi^i$.
In~(\ref{Weyl-multiplet-linearized}) we suppressed their conjugate~$\psi_{\mu}{}_i = P_R \psi_{\mu i}$ and~$\chi_i=P_R\chi_i$.
The partition function $Z^\mathcal{I}_\text{SUSY}[\widetilde\BS^4_n]$ on the branched sphere~\eqref{Branched_Sphere} can be expanded around $n=1$ as 
\begin{align}\label{LogZ_expansion}
	\begin{aligned}
& - {\rm Re} \log Z^\mathcal{I}_\text{SUSY}[\widetilde\BS^4_n] + {\rm Re} \log Z^\mathcal{I}_\text{SUSY}[\BS^4]
\\
&\quad =\int_{\BS^4} {\rm d}^4x \sqrt{g} \, \Big \langle	 \frac{1}{2}\,  \delta g_{\mu\nu}  T^{\mu\nu} +\delta\bar \psi{}_{\mu}{}^i  S^\mu{}_i +\delta\bar \psi{}_{\mu}{}_i  S^{\mu}{}^i  
+ \delta V_\mu{}^{ij}  j^{\mu}{}_{ij} + \delta A_\mu  j^{\mu}  \\
	&\qquad\quad
	+ \delta D\, J + \delta\bar{\chi}{}^i\, j_i + \delta\bar{\chi}{}_i\, j^i	+ \delta T^{+\mu\nu} j_{+\mu\nu}  + \delta T^{-\mu\nu} j_{-\mu\nu} 
	\Big\rangle_{\BS^4}^\text{(ICFT)}  +{\cal O}((n-1)^2) \ .
	\end{aligned}
\end{align}
Here we took the real parts on the left-hand side and assumed that they coincide, at least in the $n\rightarrow 1$ limit, with the conformal partition functions.
The variations of fermions are actually zero because the $Z^\mathcal{I}_\text{SUSY}[\widetilde\BS^4_n]$ background is bosonic.
In flat space one-point functions of operators with non-zero spin have to vanish due to the conformal symmetry~$SO(1,4)$ preserved by the interface~\cite{McAvity:1995zd,Billo:2016cpy}.
Most operators in the supercurrent multiplet transform as primary operators of definite weights under the Weyl transformation from flat space to a sphere, so their vevs should vanish on $\BS^4$ as well.
An exception is the stress tensor whose one-point function has a non-vanishing contribution from the conformal anomaly on a 4-sphere as in~(\ref{VEV_stress_tensor}); this case is discussed in Sections~\ref{sec:IE-sphere} and~\ref{sec:anomaly-interface}.
Assuming that the localized part of $\mathcal{A}_{\mu\nu}$ in~(\ref{VEV_stress_tensor}) does not contribute to the interface entropy,
the only non-trivial contribution from the couplings in \eqref{LogZ_expansion} comes from the scalar one-point function $\langle\, J\,\rangle^\text{(ICFT)}_{\BS^4}$ in the second line. 

We now show that, for a half-BPS superconformal interface, $\langle J\,\rangle^\text{(ICFT)}_{\mathbb{S}^{4}}$ vanishes and gives no contribution to \eqref{LogZ_expansion}.
The SUSY transformation of $j_{i}$ in flat space is given by%
\footnote{%
The components of the supercurrent multiplet for an abelian vector multiplet can be obtained by linearizing the Weyl multiplet in the superconformal action~(20.89) of~\cite{Freedman:2012zz}. Explicitly, they are given by
$T^{\mu\nu}=8 \partial^{(\mu} \bar{X}^{} \partial^{\nu)}X^{}-4g^{\mu\nu}\abs{\del_{\rho}X}^{2}
+\frac{4}{3}(g^{\mu\nu}\del^{2}-\del^{\mu}\del^{\nu})\abs{X}^{2}
-g^{\mu\nu}(\bar{X}\del^{2}X+X\del^{2}\bar{X})
+ \bar{\Omega}{}^{i} \gamma^{(\mu}\overset{\leftrightarrow}{\partial}{}^{\nu)} \Omega_{i}^{} 
-\frac{1}{4}g^{\mu\nu}\bar{\Omega}^{i}\overset{\leftrightarrow}{\sla{D}}\Omega_{i}
+2{F^{\mu}}_{\rho}F^{\nu\rho}-\frac{1}{2}g^{\mu\nu}F_{\rho\sigma}F^{\rho\sigma}$, 
$
S^\mu_{i}=-\frac{1}{2}F_{\rho\sigma} \gamma^{\rho\sigma} \gamma^{\mu} \varepsilon_{i j} \Omega^{j}
 - 2\bar{X}\overset{\leftrightarrow}{\partial}^{\mu} \Omega_{i}+2\bar{X}\gamma^\mu\slashed{\partial}\Omega_i-\frac{2}{3}  \gamma^{ \mu\nu}\partial_{\nu}\left(\bar{X} \Omega_{i}\right)
$,
$ j_{\mu}{}^{i}{}_{j}=-2\bar{\Omega}{}^{i} \gamma_{\mu}\Omega_{j}^{}+\delta^{i}_{j}\bar{\Omega}{}^{k} \gamma_{\mu}\Omega_{k}^{}$, 
$  j_{\mu}^{}=-4 \i \bar{X}^{} \overset{\leftrightarrow}{\partial}_{\mu} X^{}+\i{\bar{\Omega}}{}^{i} \gamma_{\mu}\Omega_{i}^{}$, 
$ J=-4\bar{X}X$, 
$ j_i=4\bar{X}\Omega_{i}$, 
$   j^{+}_{\mu\nu}=XF^{+}_{\mu\nu}$, and
$   j^{-}_{\mu\nu}=\bar{X}F^{-}_{\mu\nu}$ .
Here $\mathcal{L}= -4 \partial_\mu X \partial^\mu\bar X 
+\frac12\varepsilon^{ik}\varepsilon^{jl}Y_{ij}Y_{kl} 
-2\bar\Omega_i \sla{\partial} \Omega^i 
- \frac12 F_{\mu\nu}F^{\mu\nu}$ and $f\overset{\leftrightarrow}{\partial}_{\mu}  g= f \partial_\mu g - (\partial_\mu f)g$.
See also~\cite{Fisher:1982fu}.
One can obtain~(\ref{eq:susy-transformation-j}) and the other transformations from these expressions.
}
\begin{align}\label{eq:susy-transformation-j}
\delta j_{i}
=
-\frac{1}{2}(\sla{\partial}J)\epsilon_{i}
+\frac{1}{ 2}j_{\mu}{}_{i}{}^{j}\gamma^{\mu}\e_{j}
+\frac{\mathrm{i}}{2}j_{\mu}\gamma^{\mu}\e_{i}
+j_{\mu\nu}^{-}\gamma^{\mu\nu}\ve_{ij}\e^{j}.
\end{align}
As we explained earlier, the one-point functions of non-scalar operators in an ICFT vanish thanks to conformal symmetry~\cite{Billo:2016cpy}.
For constant SUSY parameters $\e_{i}$ and $\e^{i}$ 
parametrizing the supersymmetry preserved by the interface,
the Ward identity $\Braket{\,\delta j_{i}\,}^{(\text{ICFT})}_{\mathbb{R}^{4}}=0$ and 
the transformation \eqref{eq:susy-transformation-j} imply that
\begin{align}
0=\partial_{\mu}\Braket{\,J\,}^{(\text{ICFT})}_{\mathbb{R}^{4}}\gamma^{\mu}\epsilon_{i}\ .
\end{align}
Since $\del_{\mu}\Braket{\,J\,}^{(\text{ICFT})}_{\mathbb{R}^{4}}=0$ for $\mu\neq 3$, we have
\begin{align}\label{eq:condition1-for-j}
\Braket{\,J\,}^{(\text{ICFT})}_{\mathbb{R}^{4}}\gamma^{3}\epsilon_{i}=\text{constant}\ .
\end{align}
On the other hand, because $J$ has Weyl weight 2
\begin{align}\label{eq:condition2-for-j}
\Braket{\,J\,}^{(\text{ICFT})}_{\mathbb{R}^{4}}\propto\abs{y^3}^{-2}  \ .
\end{align}
For a non-zero $\epsilon_{i}$ \eqref{eq:condition1-for-j} and \eqref{eq:condition2-for-j} are compatible only if $\Braket{\,J(y)\,}^{(\text{ICFT})}_{\mathbb{R}^{4}}=0$.
Since $J$ is a conformal primary, we conclude that~$\langle\, J\,\rangle^\text{(ICFT)}_{\mathbb{S}^{4}}=0$.

Therefore, we have
\begin{align}
\log\, Z^\mathcal{I}_\text{SUSY}[\BS^4_n] =\log\, Z^\mathcal{I}_\text{SUSY}[\BS^4]+{\cal O}((n-1)^2)\ ,
\end{align}
for the supersymmetric R\'enyi entropy. 
This shows that the second term in~(\ref{ICFT_SRE-derivative}) vanishes.

\bibliographystyle{JHEP}

\bibliography{Interface_EE2}

\end{document}